\documentclass[11pt,a4paper]{article}
\usepackage{jcappub}

\pdfoutput=1 

\usepackage[T1]{fontenc} 
\usepackage[utf8]{inputenc}
\usepackage{upgreek}
\usepackage{textgreek}
\usepackage{amsmath}
\usepackage{amssymb}
\usepackage{graphicx}
\usepackage{dcolumn}
\usepackage{multirow}
\usepackage{bm}
\usepackage[dvipsnames]{xcolor}
\usepackage[normalem]{ulem}
\usepackage{xspace}
\usepackage{aas_macros}
\usepackage{csquotes}
\usepackage{threeparttable}
\usepackage{tabularx}
\usepackage{soul}
\usepackage{placeins}

\usepackage{adjustbox}

\usepackage{todonotes}
\usepackage[capitalize]{cleveref}

\newcommand{\final}[1]{{#1}}
\newcommand{\finalr}[2]{{#2}}

\newcommand{\Mpc}{\mathrm{Mpc}}
\newcommand{\Mpcinv}{\mathrm{Mpc}^{-1}}

\newcommand{\hinvMpc}{\mathrm{Mpc}/h}
\newcommand{\hinvkpc}{h^{-1}\mathrm{kpc}}

\newcommand{\lya}{{\ensuremath{\mathrm{Ly}\alpha}}} 

\newcommand{\skm}{\mathrm{s \, km^{-1}}}
\newcommand{\gadget}{\texttt{gadget}\xspace}
\newcommand{\Borde}{B13\xspace}

\newcommand{\HeII}{He\,\textsc{ii}}
\newcommand{\SiII}{Si\,\textsc{ii}}
\newcommand{\SiIII}{Si\,\textsc{iii}}

\newcommand{\PoneD}{$P_\mathrm{F,1d}$\xspace}

\newcommand{\gridname}{Lyssa}
\newcommand{\gridlongname}{Lyman-$\alpha$ simulation suite in absorption}

\begin{document}

\hfill \texttt{TTK-24-54}\\

\title{\boldmath Emulating the Lyman-Alpha forest 1D power spectrum from cosmological simulations: New models and constraints from the eBOSS measurement}

\author[a,b,1]{Michael Walther\note{corresponding authors}}
\author[a,b,1]{Nils Sch\"oneberg}
\author[c]{Sol\`ene Chabanier}
\author[d]{Eric Armengaud}
\author[c]{Jean Sexton}
\author[d]{Christophe Y\`eche}
\author[e]{Julien Lesgourgues}
\author[e,f]{Markus R. Mosbech}
\author[g,d]{Corentin Ravoux}
\author[c,d]{Nathalie Palanque-Delabrouille}
\author[c]{Zarija Luki\'c}

\affiliation[a]{University Observatory, Faculty of Physics, Ludwig-Maximilians-Universität, Scheinerstr. 1, 81677 Munich, Germany}
\affiliation[b]{Excellence Cluster ORIGINS, Boltzmannstr. 2, 85748 Garching, Germany}
\affiliation[c]{Lawrence Berkeley National Laboratory, 1 Cyclotron Rd, Berkeley, CA 94720, USA}
\affiliation[d]{Universit\'e Paris-Saclay, CEA, IRFU, 91191, Gif-sur-Yvette, France}
\affiliation[e]{Institute for Theoretical Particle Physics and Cosmology (TTK), RWTH Aachen University, 52056 Aachen, Germany}
\affiliation[f]{Institute for Theoretical Particle Physics (TTP), Karlsruhe Institute of Technology (KIT), 76128 Karlsruhe, Germany}
\affiliation[g]{Université Clermont-Auvergne, CNRS, LPCA, 63000 Clermont-Ferrand, France}
\emailAdd{michael.walther@lmu.de}
\emailAdd{nils.science@gmail.com}


\abstract{
We present the \gridname{} suite of high-resolution cosmological simulations of the Lyman-$\alpha$ forest designed for cosmological analyses. These 18 simulations have been run using the \texttt{Nyx} code with $4096^3$ hydrodynamical cells in a 120 Mpc ($\sim 81\, \hinvMpc$) comoving box and individually provide sub-percent level convergence of the Lyman-$\alpha$ forest 1d flux power spectrum. We build a Gaussian process emulator for the \gridname{} simulations in the \texttt{lym1d} likelihood framework to interpolate the power spectrum at arbitrary parameter values. We validate this emulator based on leave-one-out tests and based on the parameter constraints for simulations outside of the training set. We also perform comparisons with a previous emulator, showing a percent level accuracy and a good recovery of the expected cosmological parameters. 
Using this emulator we derive constraints on the linear matter power spectrum amplitude and slope parameters $A_\lya$ and $n_\lya$\,. While the best-fit Planck $\Lambda$CDM model has $A_\lya=8.79$ and $n_\lya=-2.363$, from DR14 eBOSS data we find that \mbox{$A_\lya<7.6$} \mbox{(95\% CI)} and \mbox{$n_\lya=-2.369 \pm 0.008$}. The low value of $A_\lya$\,, in tension with Planck, is driven by the correlation of this parameter with the mean transmission of the Lyman-$\alpha$ forest. This tension disappears when imposing a well-motivated external prior on this mean transmission, in which case we find $A_\lya=9.8\pm1.1$ in accordance with Planck.
}



\maketitle

\section{Introduction}
The Lyman-$\alpha$ forest consists of a densely spaced series of absorption features in the spectra of distant quasars. These features track the distribution of neutral hydrogen and can consequently be used to probe cosmological structures. As the Lyman-$\alpha$ forest is sensitive to high redshift ($2<z<5$) structures at small scales up to around $k_\mathrm{max} \sim 5/\mathrm{Mpc}$, it probes structure formation at the smallest scales in the mildly nonlinear regime. However, the Lyman-$\alpha$ forest is also very sensitive to the thermal state of the intergalactic medium (IGM), such as its mean temperature or the pressure smoothing scale, and thus computationally expensive hydrodynamical simulations are typically required to model this observable with high accuracy. For a review, see \cite{Meiksin+09,McQuinn+15}.

The 1d power spectrum of the Lyman-$\alpha$ forest (\PoneD{}) is the power spectrum of the so-called transmission contrast $\delta_F=F/\bar{F}-1$ where $F$ is the flux transmission of the Lyman-$\alpha$ forest and $\bar{F}$ is its mean. Recent interest for \PoneD{} has grown due to it being an excellent probe of many models that predict a small-scale suppression or enhancement of power, as well as from claims of a high tension between the power measured with the Lyman-$\alpha$ forest and that expected from the CMB anisotropies \cite{Palanque+19,Rogers+23,Goldstein+23} (see also below). In particular, due to the high reach in wavenumbers and redshift, high precision measurements of the \PoneD{} can be used to put precise constraints on many dark matter candidates (warm dark matter, interacting dark matter, fuzzy dark matter), neutrino masses, inflationary scenarios predicting a running of the power spectrum, and other models that impact the small-scale power spectrum \cite{Palanque+19, Rogers+23, Goldstein+23}.

The Lyman-$\alpha$ forest has been measured to high precision by the extended Baryon Oscillation Sky Survey (eBOSS), which is part of the Sloan Digital Sky Survey, using data~release~14 (SDSS~DR14), where around 44\,000 quasars provide a near-percent level statistical error on \PoneD~\cite{Chabanier+19}.
Furthermore, the ongoing Dark Energy Spectroscopic Instrument (DESI) survey~\cite{DESI+23} will also measure \PoneD{} at excellent precision, with improved statistics and spectrograph resolution~\cite{DESI+22}. Additionally, there are various datasets containing smaller amounts of high-resolution quasar spectra allowing to extend the survey measurements to smaller scales, albeit at lower precision (see \cite{Karacayli+21}). These also require a high degree of accuracy in the theoretical predictions. As such, there is a great need for highly converged, accurate simulations of the Lyman-$\alpha$ forest.

\enlargethispage*{2\baselineskip}
There have been a number of recent approaches to simulating \PoneD{} with different simulation tools and parametrizations, which we summarize in \cref{tab:simulation_overview}. Notably, about a decade ago the authors of \cite{Borde+13} have created a suite of simulations for the purpose of analyzing  BOSS data based on the \gadget code (hereafter \Borde), splicing together low- and high-resolution simulations (see \cref{tab:simulation_overview}). The latest analysis of eBOSS data with this set of simulations found a tension in the cosmological parameters compared to those derived from CMB anisotropies within the $\Lambda$CDM framework. In the original work by~\cite{Palanque+19}, based on the full likelihood modeling of the Lyman-$\alpha$ data, this tension was found at the 3.6$\sigma$ level. More recent studies, focusing on the slope and amplitude of the matter power spectrum at the Lyman-$\alpha$ scale, made use of the likelihood contours in this reduced parameter space (previously published in~\cite{Chabanier+19}), and found a higher tension at the level of 4.9$\sigma$~\cite{Rogers+23, Goldstein+23}.

Recent simulation suites listed in \cref{tab:simulation_overview} have typically not been applied to eBOSS data (due to insufficient resolution, lack of convergence, or not varying cosmological parameters), with the exception of the PRIYA suite described in \cite{Bird+23,Fernandez+23}. This suite involves a number of low- and high-resolution simulations that are used to build a multi-fidelity emulator.


In this paper, we present a new suite of simulations, which we dub \enquote{\gridlongname} (\gridname), designed to provide high precision predictions for the one-dimensional flux power spectrum of the Lyman-$\alpha$ forest at both small and large scales with a box of size $L=$120~Mpc and resolution $29.3\,\mathrm{kpc}$, defined as $L/N^{1/3}$ where $N$ is the total particle number. As such, each simulation is covering modes from $k_\mathrm{min}=0.052\,\Mpcinv$ ($=2\pi/L$) to the Nyquist frequency $k_\mathrm{Nyq}=107\,\Mpcinv$. However the actual maximum wavenumber used in this work is $k_\mathrm{max}=5\,\Mpcinv$, to ensure proper convergence to the sub-$\%$ level in the wavenumber range corresponding to current Lyman-$\alpha$ forest data.
Using these simulations, we build an emulator that can be used to obtain constraints on the small scale power spectrum amplitude and slope. This information can then, in turn, be compared to the predictions of various cosmological models, in particular those claiming to ease the $\sigma_8$ tension (see for example~\cite{Hooper:2021rjc,Hooper:2022byl}). This new simulation suite is targeted for analyses of the \PoneD{} in eBOSS and DESI, though it might also be useful in future analysis of the three-dimensional flux transmission power spectrum.

\finalr{Our new simulation suite is roughly comparable to the high-fidelity simulations of \cite{Bird+23} (18 compared to 3 simulations, though without the 48 low fidelity simulations), although using a much smaller box and with about twice the resolution at mean density}{}\final{Comparing the \gridname{} simulations to the high-fidelity simulations of \cite{Bird+23}, we see that \gridname{} has 18 compared to 3 high-resolution simulations. Instead, Lyssa is missing the 48 low fidelity simulations (which are used in \cite{Bird+23} to construct a multi-fidelity emulator) and therefore covers cosmological parameter space with a lower density. Compared to the high-fidelity simulations, Lyssa has a smaller box, but higher resolution at mean density}. However, the resolution at mean density does not necessarily provide a complete description of convergence properties. For example, \cite{Chabanier+21} compared Nyx to CRK-HACC showing that Nyx reaches similar \PoneD{} accuracy with half the resolution at mean density at $z=3$ if SPH-based density estimation is used for CRK-HACC (and better accuracy at $z=5$ under the same conditions). Given the smaller box size, we caution about cosmic variance concerns, see \cite{Borde+13,Bird+23,Pedersen+19}. More detailed comparisons of the fidelity of the simulations between \cite{Bird+23}, \cite{Chabanier+21}, and this work as well as more detailed convergence studies overall are left for future work. For resolution convergence tests of our setup based on smaller simulation boxes, see \cite{Walther+20}.


\begin{table}
    \centering
\begin{threeparttable}
    \begin{tabular}{@{}c@{}}
    \resizebox{\textwidth}{!}{
    \begin{tabular}{l| r r | r r | c}
        Simulation suite & \multicolumn{2}{c |}{Box size} &  \multicolumn{2}{c|}{Resolution} & Code \\ \hline
        \multirow{3}{*}{\Borde{}$^{*}$ \cite{Borde+13}} & 25$\,\hinvMpc$ &$\sim\,$37.0\,Mpc & $130.2\,\hinvkpc$ & $\sim\,$193\,kpc & \multirow{3}{*}{GADGET-3}\\ 
        & 100$\,\hinvMpc$&$\sim\,$148\,Mpc & $130.2\,\hinvkpc$ & $\sim\,$193\,kpc \\
        & 100$\,\hinvMpc$&$\sim\,$148\,Mpc & 32.55$\,\hinvkpc$ & $\sim\,$48.2\,kpc\\ \hline
        B19$^{*}$ \cite{Bird+19} & 40$\,\hinvMpc$ & $\sim\,$57.1\,Mpc & 156.3$\,\hinvkpc$& $\sim\,$223\,kpc & MP-GADGET\\ \hline
        THERMAL \cite{Walther+19} & 20$\,\hinvMpc$ & $\sim\,$29.8\,Mpc & 19.5$\,\hinvkpc$ & $\sim\,$29.1\,kpc & Nyx \\ \hline
        P21$^{*}$ \cite{Pedersen+20,Pedersen+23,Cabayol+23} & $\sim\,$47.3$\,\hinvMpc$ & 67.5\,Mpc &$\sim\,$61.4$\,\hinvkpc$ &87.7\,kpc & MP-GADGET\\ \hline
        SHERWOOD$^{+}$ \cite{Bolton+16,Puchwein+22} & 40$\,\hinvMpc$ & $\sim\,$59.0\,Mpc & 19.5$\,\hinvkpc$ & $\sim\,$28.8\,kpc & P-Gadget3\\ \hline
        \multirow{2}{*}{PRIYA \cite{Bird+23,Fernandez+23}} & 120$\,\hinvMpc$ &$\sim\,$171\,Mpc  & 39.1$\,\hinvkpc$ & $\sim\,$55.9\,kpc  & \multirow{2}{*}{MP-GADGET}\\
        & 120$\,\hinvMpc$ &$\sim\,$171\,Mpc  & 78.1$\,\hinvkpc$ & $\sim\,$111\,kpc \\ \hline 
        \gridname~(this work) & $\sim\,$80.8$\,\hinvMpc$ & 120\,Mpc & $\sim\,$19.7$\,\hinvkpc$ & 29.3\,kpc & Nyx
    \end{tabular}
    }
    \end{tabular}
    \caption{Summary of different simulation suites for \PoneD Lyman-$\alpha$ analyses. The conversions have been performed with the given central values of $h$ or the mean of the ranges, and are marked with \enquote{$\sim$} as different simulations do not necessarily share the same box size in those units. Note that for SPH codes, mass resolution is fixed and the \enquote{Resolution} column shows the value 
    for gas at mean density. In over-/underdense regions the SPH resolution is accordingly better/worse.
    }
    \label{tab:simulation_overview}
    \begin{tablenotes}\footnotesize
        \item[*] These simulation suites are not named. They are described in detail in references \cite{Borde+13} (\Borde), \cite{Bird+19} (B19) and \cite{Pedersen+20} (P21), respectively. Note that the \Borde{} suite has been used to generate the Taylor emulator we compare to in this work.
        \item[+] In this case we are only quoting the simulation with the best resolution among those with the highest number of particles.
    \end{tablenotes}
\end{threeparttable}
\end{table}

\enlargethispage*{1\baselineskip}
This paper is structured as follows. In \cref{sec:sims} we describe the new high-precision suite of simulations \enquote{\gridname{}}, the choice of the thermal and cosmological parameter basis, as well as the overall likelihood pipeline. In \cref{sec:validation} we validate the emulator using leave-one-out tests and cosmological parameter fits. In \cref{sec:simulation-comp} we compare with simulations from \Borde{} and check the agreement with the results~of~\cite{Palanque+19}. In \cref{sec:data} we then obtain constraints from the eBOSS data using the emulator, and we conclude in \cref{sec:conclusion}.

\section{\gridname{} simulations and inference pipeline}\label{sec:sims}
In this section, we describe the set of simulations and the corresponding emulator that has been built in preparation for the upcoming DESI data. We highlight the similarities and differences of \gridname{} with respect to other similar simulation suites, and in particular with the simulation set from~\cite{Borde+13} for which an emulator has been designed based on a Taylor approximation around a central simulation. We refer to this emulator as the \emph{Taylor emulator}.

\subsection{Individual simulation design}\label{ssec:individual_sim}

\begin{figure}
\centering\includegraphics[height=0.4\textheight]{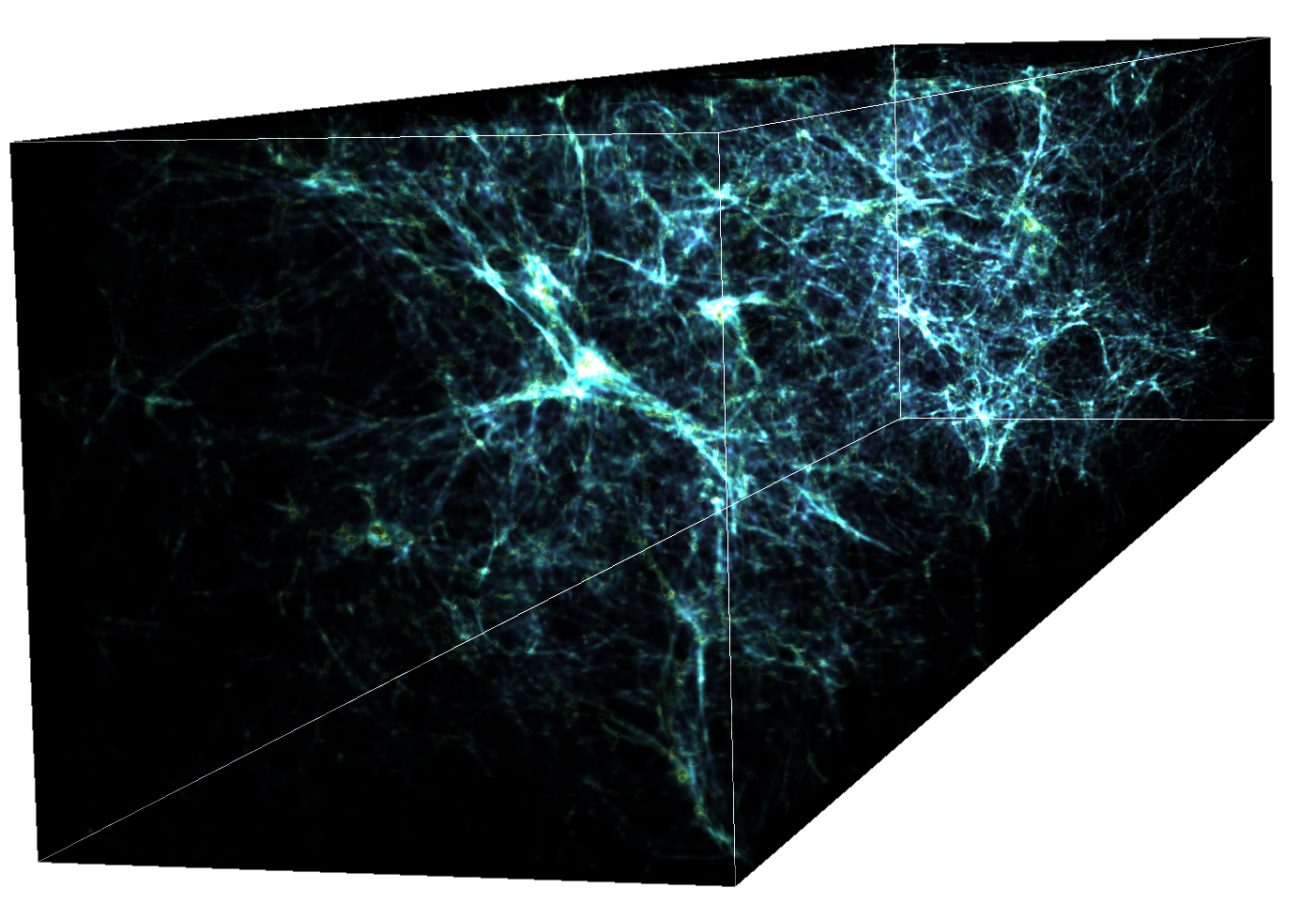}
\caption{A rendering of a $21\,\mathrm{Mpc}\times 21\,\mathrm{Mpc}\times 120\,\mathrm{Mpc}$ cutout from one of the simulation boxes. Brighter colors correspond to larger baryon density.}
\label{fig:rendering}
\end{figure}
To generate an individual model of the Lyman-$\alpha$ forest power spectrum, we proceed as follows. First, we run a cosmological hydrodynamical simulation with the \texttt{Nyx} code\footnote{\url{https://github.com/AMReX-Astro/Nyx}} \citep{Almgren+13,Lukic+15,Sexton2021}. \texttt{Nyx} evolves a hydrodynamical fluid on an Eulerian grid, which is gravitationally coupled to N\nobreakdash-body particles tracing the dark matter, all deposited on the same grid. While adaptive mesh refinement techniques are available in \texttt{Nyx}, they are inefficient for Lyman-$\alpha$ forest simulations as the gas of interest spans most of the simulation volume. The Euler equations are solved with a second\nobreakdash-order accurate scheme, and key physical processes affecting the intergalactic medium -- such as photoionization, collisional ionization, recombination, heating, and atomic cooling -- are included in the simulation. We show an example cutout of the simulation box in \cref{fig:rendering}.

Based on the convergence considerations of \cite{Lukic+15,Walther+20,Chabanier+21}, each simulation run is using $4096^3$ hydrodynamical cells (plus an equal amount of dark matter particles) in a $120\mathrm{Mpc}$ comoving~box.\footnote{Note that the \gridname{} fiducial simulation is in fact the largest simulation run in \cite{Walther+20} (labeled $120\,\mathrm{Mpc}~4096^3$). Both resolution and box size convergence have been studied there. For an internal cross-check on cosmic variance, we compared extractions taken along the different simulation axes. Further tests using initial conditions with different seeds are planned for future work.} Notably, the simulations cover the whole range of scales probed by the \PoneD{} measurement of DESI, without resorting to the splicing method used in \cite{Borde+13} for the Taylor emulator. We initialize the simulations at $z_{\mathrm{ini}}=99$ and take snapshots of the evolved fluid properties and dark matter densities at thirteen redshifts in intervals of $0.2$ between $z=2.2$ and $4.6$, as well as two additional high-redshift snapshots at redshifts $5.0$ and $5.4$. 

To generate initial conditions for our simulations, we use \texttt{2lpt-ic} \cite{Crocce+12,Crocce+2012_software} and produce particles following the total matter transfer function at $z_{\mathrm{ini}}$ (see \cref{ssec:ic} for a comparison with the individual transfer functions for both fluids and a different IC code: \texttt{MUSIC2-monofonIC}). The baryonic part of the particle mass is then subtracted from the particles and deposited on a regular mesh using a cloud\nobreakdash-in\nobreakdash-cell scheme.

\enlargethispage*{2\baselineskip}
The effect of a time-varying, spatially uniform ultraviolet background (UVB) on photo-ionization and photo-heating is modeled by the code based on a list of redshift-dependent rates~\citep{Katz+92}. These rates were derived from the late reionization model of~\cite{Onorbe+18}, and a warm \HeII{} reionization scenario with $\Delta T=20000K$, whose resulting thermal evolution is in good agreement with the measurement of~\cite{Walther+19}. To allow different thermal evolutions, we rescale the overall heating rates by an additional density- and redshift-independent factor $A_\mathrm{UVB}$\,. Considering two simulations with different $A_\mathrm{UVB}$\,, even after rescaling their snapshot's temperature at a given redshift to a common value (as described below), they will still have different gas pressure smoothing scales $\lambda_P$ which depends on the full thermal history of the gas within the simulation (see for example~\citep{Kulkarni+15}). Therefore, within this approach, the $A_\mathrm{UVB}$ parameter allows us to vary the smoothing scale. By interpolating between simulations with different $A_\mathrm{UVB}$ values for different redshifts, an arbitrary $\lambda_P(z)$ history can be emulated.

After the simulations have been run, we generate Lyman-$\alpha$ forest spectra from the snapshots and compute 1d- and 3d- power spectra of different quantities (baryons, dark matter, matter, Lyman-$\alpha$ forest flux transmission), making use of the \texttt{gimlet} postprocessing suite~\cite{Friesen+16}. This software package allows for efficient, MPI-parallel computation of typical statistical properties in a \texttt{Nyx} simulation based on the hydrodynamical grid and cloud\nobreakdash-in\nobreakdash-cell depositions of the dark matter particles.  The extraction of flux skewers is performed along each of the principal axes of the box and statistical properties are averaged over all 3 axes to reduce cosmic variance. 
As described in \cite{Walther+20}, to allow generating a variety of thermal states at each redshift we additionally rescale the output temperatures in all pixels by a power law in $(1+\delta)=\rho_b/\bar{\rho}_b$, where $\rho_b$ is the baryon density and $\bar{\rho}_b$ is its spatial average. This allows us to emulate, with arbitrary redshift dependence, the coefficients $T_0$ and $\gamma$ of the usual temperature-density relation
\begin{equation}\label{eq:tdr}
    T(\delta,z) = T_0(z) (1+\delta)^{\gamma(z)-1}
\end{equation}
Explicitly, we do not set all pixel temperatures equal to the power law, but instead we keep the scatter of the temperature-density relation intact. According to \cite{takhtaganov2019cosmic} this allows for computing \PoneD~with sub-percent accuracy for $k < 0.03 \,\skm$ (at $z=3.0$) compared to running a new simulation with a rescaled thermal history. Finally, to match a range of mean transmissions, we rescale the resulting optical depths with a global factor $f_{\tau}$. 

We evaluate $\lambda_P$ by fitting a cutoff power law $A k^n \exp(-k^2 \lambda_P^2)$ to the \enquote{real space} three-dimensional flux power spectrum, which is estimated similarly to the (redshift-space) flux power spectrum, but ignoring line broadening and peculiar velocity shifts during skewer generation. When computing the real space flux field, we rescale the real space optical depth of the simulation ($\tau_\mathrm {real}$) by the same factor $f_\tau$ which is needed for the redshift space transmission to match the fiducial  mean transmission ($\bar{F}$) for a given redshift. 

We acknowledge that some physical effects are not included in our simulation suite. These will be included as additional corrections during the likelihood fit, see \cref{ssec:nuisances}. While we already account for SN and AGN feedback in this work (similar to the previous eBOSS analysis by \cite{Palanque+19}), further corrections of e.g. inhomogeneous \HeII-reionization are left for future work. Therefore, the constraints on \finalr{$n_\mathrm{\lya}$ }{}\final{cosmological parameters} shown in \cref{sec:data} might \finalr{be tighter than with }{}\final{change once} these physical effects \final{are} marginalized over (see e.g. \cite{Onorbe+19,Bird+23, Upton-Sanderbeck+20} for the effects of \HeII{} reionization on the \PoneD).

\subsection{Cosmological parameterization and suite design}\label{ssec:design}
For the purpose of the Lyman-$\alpha$ \PoneD{} analysis, we would ideally sample simulation parameters from a broad region based on the (expected) constraints of current (and upcoming) data. 

Some simulation suites dedicated to the cosmological analysis of the flux power spectrum, including in particular the \Borde suite and its Taylor emulator \cite{Borde+13, Palanque+19} have used cosmological base parameters widely used within the cosmological community (especially $n_s$ and $A_s$/$\sigma_8$), which were commonly used in CMB anisotropy measurements and low-redshift galaxy surveys. However, these cannot be connected in a straightforward way to the Lyman\nobreakdash-$\alpha$ forest data probing both smaller scales than the CMB and higher redshifts than most galaxy surveys.

Other recent works such as \cite{Bird+19, Pedersen+23} have adopted a different basis of cosmological parameters, more adapted to the actual sensitivity of \PoneD{} measurements. In particular, as already found by~\cite{McDonald+05c:arxiv0407377}, \PoneD{} best constrains the amplitude and slope of the linear matter power spectrum computed at the redshift and pivot wavenumber of the Lyman-$\alpha$ data.
Following this approach, we define the parameters $A_\lya$ and $n_\lya$ as the amplitude and logarithmic slope of the linear matter power spectrum $P_\mathrm{lin}(k_{\rm p},z_{\rm p})$ at a pivot redshift of ${z_{\rm p}=3}$ and a pivot scale of ${k_{\rm p}=1\,\Mpcinv}$.\footnote{In practice we derive these parameters by convolving the power spectrum with a narrow Gaussian of width $\Delta \ln k=0.1$ (for the slope we use the derivative of a Gaussian). This allows us to reduce numerical errors from interpolation. We have checked that the results are consistent with similar definitions. Whenever we show $A_\lya$ it is implicitly assumed to be in units of $\mathrm{Mpc}^3$.}
For comparison, \cite{Pedersen+23} makes use of the parameters $\Delta_\star$ and $n_\star$ at $k_{\rm p}=0.009\,\skm{}$ (about 0.7Mpc${}^{-1}$), while~\cite{Bird+23} defines $A_{\rm p}$ as the amplitude and $n_{\rm p}$ as the slope of the primordial power spectrum at $k_{\rm p}=0.78\,\Mpcinv$. Note that conversions between $\Delta_\star^2, n_\star$ and $A_\lya, n_\lya$ can be approximated as 
\begin{align}
    \Delta_\star^2 & \approx A_\lya \, \frac{k_\star^3}{2\pi^2} \frac{H(z)}{(1+z)} \, \left[\frac{k_\star}{k_\lya}\right]^{n_\lya}~, \\
    n_\star & \approx n_\lya~
\end{align}
or otherwise converted, for example by building an emulation scheme for their mapping (which we employ in \cref{fig:moneyplot} and \cref{tab:results}).

Within the flat $\Lambda$CDM model, the other parameters we use for the suite are the physical matter density $\Omega_{\rm m} h^2$ and $h$ (defined as $H_0 = h \cdot 100\mathrm{km/s/Mpc}$). Indeed, as highlighted in~\cite{Pedersen+19,Pedersen+23}, the sensitivity of \PoneD{} measurements to $\Omega_{\rm m}$ and $H_0$ is driven, at first order, by the conversion factor between inverse velocity units (s km$^{-1}$) from spectroscopic observations and physical units in $\Mpc^{-1}$, which is:
\begin{equation}
    \frac{H(z)}{1+z} = \frac{100\,\mathrm{km/s/Mpc}}{1+z} \, \sqrt{\Omega_{\rm m} h^2 [(1+z)^3-1] +h^2}
\end{equation}
At redshift $z>2$, the effect of dark energy on the Hubble rate is subdominant, so that to a good approximation this conversion factor scales like $\sqrt{\Omega_{\rm m} h^2 \, (1+z)}$. To avoid artificial parameter dependencies, we therefore use $\Omega_{\rm m} h^2$ as a design parameter. While our suite also varies $h$ as a cosmological parameter, we confirm in the following section that, as expected according to~\cite{Pedersen+19,Pedersen+23}, realistic \PoneD{} measurements are not expected to provide a significant sensitivity to this additional parameter.

We explicitly keep $\Omega_b h^2$ fixed to $0.02233$ based on \citep{Planck:2018vyg} and our simulations are run with massless neutrinos. As a matter of fact, the flux power spectrum measured from simulations with either massive or massless neutrinos are nearly indistinguishable from each other provided that one performs a re-mapping of redshifts within the massless neutrino simulation such that the linear matter power spectrum amplitude evolution $A(z)$ is equivalent to that of the massive neutrino simulation. Therefore, the sum of neutrino masses can still be constrained using this simulation suite, as was demonstrated in \cite{Pedersen+19,Pedersen+23}. To summarize, in total our simulations are characterized by a set of five parameters $\{A_\lya\,, n_\lya\,, \Omega_{\rm m} h^2, h, A_\mathrm{UVB}\}$. The UV background scaling parameter $A_\mathrm{UVB}$ was discussed in \cref{ssec:individual_sim}. 

\begin{figure}
    \centering
    \includegraphics[scale=0.73]{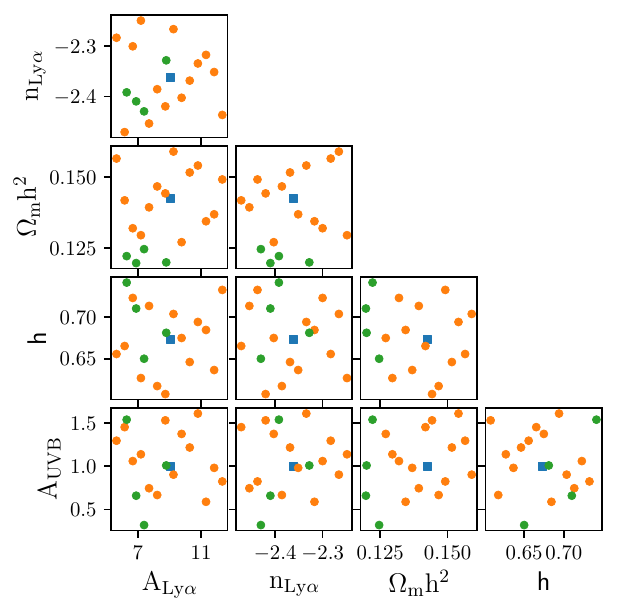}
    \includegraphics[scale=0.73]{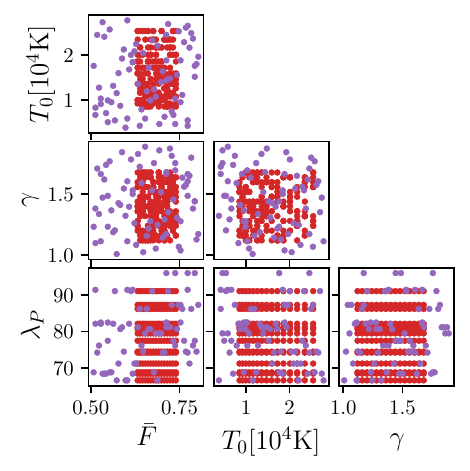}
    \caption{The parameter designs sampled by our hydrodynamical simulations. Left: Cosmological parameter space. Here the orange points are from the base latin hypercube design, the green points are from the refinement simulations and the blue point is from the fiducial. Right: Thermal parameter space at $z=3$ (for the redshift evolution, see \cref{fig:therm-grid}). In purple we show the main suite, and in red the auxilliary suite. The sparser clustering of $\lambda_P$ around specific values arises from strong anti-correlation with $A_\mathrm{UVB}$ (as it is not explicitly modified in post-processing). The values for $\lambda_P$ compare well to existing measurements like~\cite{Rorai+2017}.}
    \label{fig:parameter-grid}
\end{figure}

The original layout of the suite, illustrated in~\cref{fig:parameter-grid}, is a Latin hypercube design in the varied parameters using a total of 14 simulations. It also includes a \enquote{fiducial} simulation at the approximate center of parameter ranges, which was already described in \cite{Walther+20}. For historic reasons, this simulation features a slightly different box length (see \cref{tab:sim_parameters}) and was initialized with \texttt{camb} instead of \texttt{class}. Later, we also added 4  refinement simulations to increase the coverage of lower values of $\Omega_{\rm m} h^2$\,. The cosmological parameters of each of the 18 simulations are given in \cref{tab:sim_parameters}.

\begin{figure}
    \centering
    \includegraphics[width=0.32\textwidth]{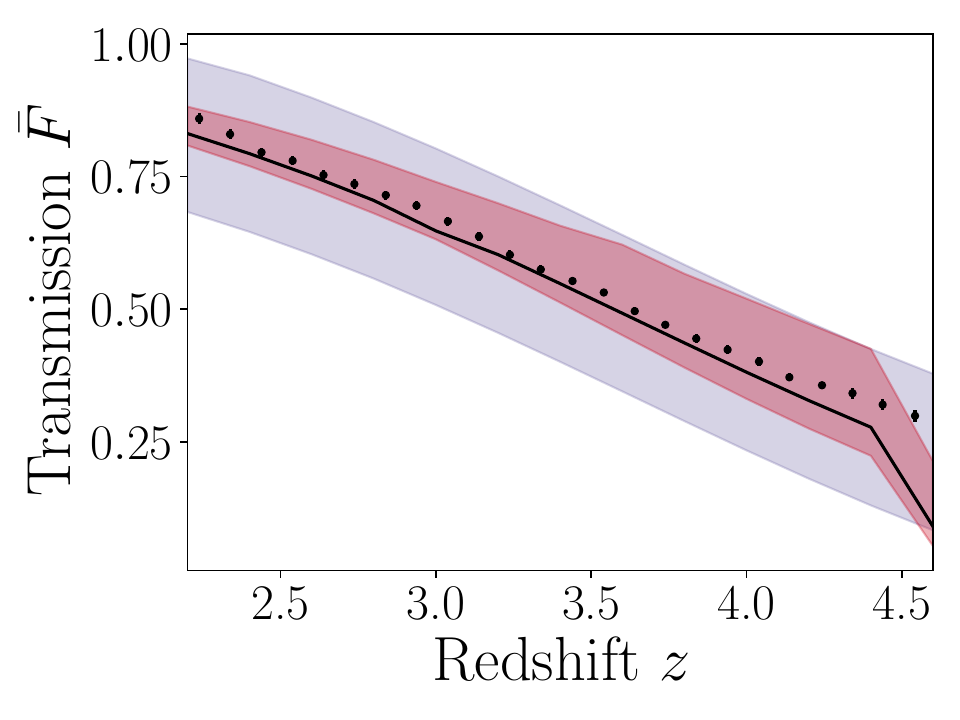}
    \includegraphics[width=0.32\textwidth]{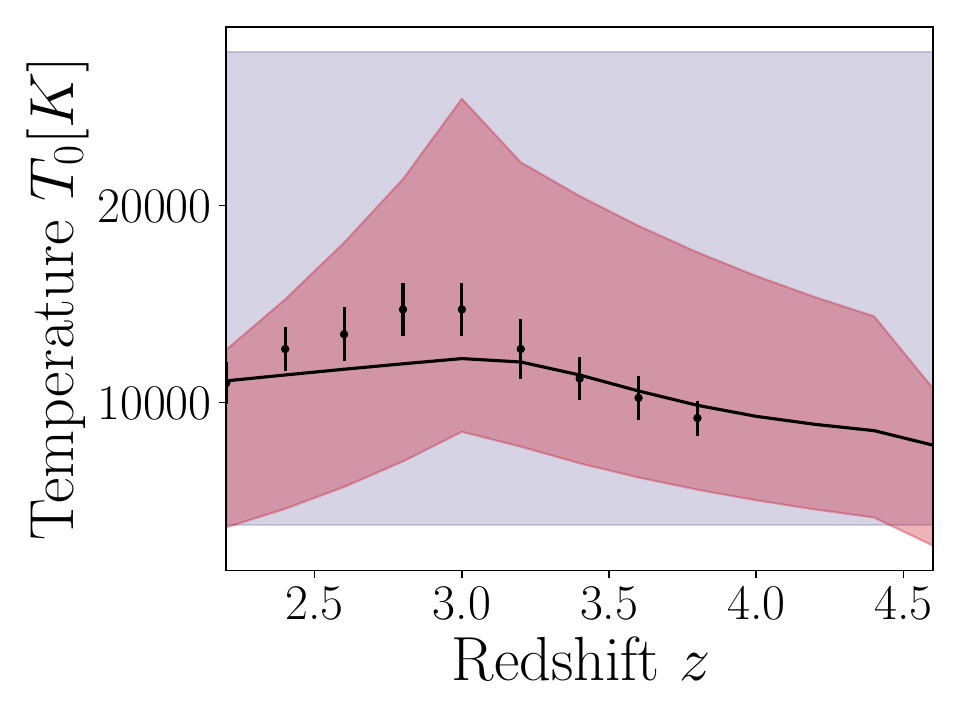}
    \includegraphics[width=0.32\textwidth]{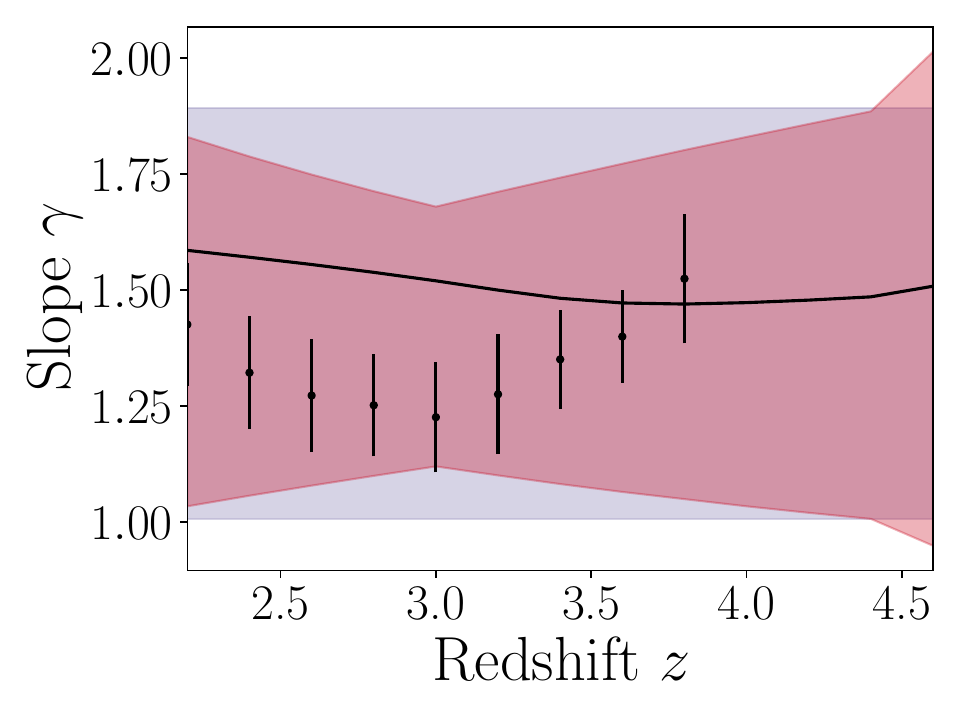}
    \caption{Redshift evolution of the thermal parameters in the different post-processings of the simulations. We show in purple/red the entire covered region (mininum and maximum covered values) of the main/auxilliary suite, and in black the fiducial thermal history. The black points represent data from \cite{Gaikwad:2020eip} ($T_0$ and $\gamma$) and \cite{Becker+13} ($\bar{F}$) just for reference. 
    }
    \label{fig:therm-grid}
\end{figure}

In addition to the parameters mentioned above, at each redshift and for each simulation, we generate a wide latin hypercube design in the parameters describing the thermal model, that is, the mean transmission $\bar{F}(z)$ and the temperature-density relation of \cref{eq:tdr}, using the approach described in \cref{ssec:individual_sim}. We use a combination of two different thermal model designs:
\begin{itemize}
    \item In the main suite, we sample the ranges $3800\mathrm{K} < T_0 < 27800 \mathrm{K}$, $1.0 < \gamma < 1.9$, and $|\bar{F}-\exp(-0.0025 (1+z)^{3.7})|<0.15$. For each of the 18 simulations we post-process 4 thermal samples for a total of 72 samples per redshift, see the purple points of \cref{fig:parameter-grid}.
    \item In an auxilliary suite we use a narrower Latin hypercube design of 15 thermal models per cosmological parameter point of the base suite ($15 \times 14 = 210$) (red points in \cref{fig:parameter-grid}) forming a Latin hypercube design for each simulation.\footnote{Those were generated by exploring the ranges covered by the 3 parameters at each redshift for bounds defined via, firstly, $\bar{F}=\exp(-0.0025 A_F (1 + z)^{B_F+3.7} (1+z_\mathrm{p})^{-B_F})$ for $0.7<A_F<1.1$ and $|B_F|<0.25$; secondly, a broken power-law in $T_0 = T_\mathrm{p} ((1+z)/(1+z_\mathrm{p}))^{\eta(z)}$ for $12000<T_\mathrm{p}/\mathrm{K}<18000$, $|\eta(z<z_\mathrm{p})-2.1|<0.05$, and $|\eta(z>z_\mathrm{p})-3.4|<0.05$; and thirdly, a power law in $\gamma=\gamma_\mathrm{p} ((1+z)/(1+z_\mathrm{p}))^{\xi}$ with $1.3<\gamma_\mathrm{p}<1.7$ and $|\xi|<0.1$. We used $z_\mathrm{p}=3$ for all parameters.}
\end{itemize}
We show the evolution of the covered thermal parameter ranges in \cref{fig:therm-grid}.

\enlargethispage*{1\baselineskip}
\subsection{Emulator}\label{ssec:emu}
We use the previously described set of simulations with a Gaussian process emulator based on the \texttt{george} package \cite{Ambikasaran+15}. The emulator code is publicly available as part of the \texttt{lym1d} package.\footnote{The \texttt{lym1d} python package is freely available at \url{https://github.com/schoeneberg/lym1d}. It includes the Gaussian Process emulator presented here, a version of the Taylor expansion used in \cite{Palanque+19}, and the likelihood framework used throughout this work. It does not include the \gridname{} simulations because of the large file size.} The emulator code builds a separate Gaussian process at each redshift that predicts all wavenumber bins of the one-dimensional flux power spectrum (in Mpc units) as a function of the cosmological and thermal parameters. 

The total set of 8 input parameters for the emulator is thus
\begin{equation}
    \{ A_\lya , n_\lya , \Omega_{\rm m} h^2, h, T_0(z),  \gamma(z), \bar{F}(z), \lambda_P(z) \}.
\end{equation}
In addition to this default set of basis parameters, a few alternative parameterizations are also implemented. First, it is possible to use $\{\Delta_*, n_*\}$ or $\{\sigma_8, n_s\}$ instead of $\{A_{\lya}, n_{\lya}\}$. Second, the native parameter $A_{\rm UVB}$ may be used in place of $\lambda_P$\,. For the Gaussian process we use a stationary Matern-5/2 covariance function with an independent length scale $l_i$ for each of the eight emulation parameters $i$ and an overall scaling $\sigma_0$, which were optimized for each wavenumber and redshift bin independently. The typical length scales we obtain fall between~3~and~20 times the range covered by simulations for each parameter. For the hyperparameter optimization, we use all thermal re-processings for all simulations, leading to a total of 297 points per redshift. More details about how the emulator is generated can be found in \cite{Walther+20}.

When using the emulator one can either specify the thermal parameters \{$T_0$, $\gamma$, $\bar{F}$, $\lambda_P$\} at each redshift independently or describe them by (broken) power laws. For simplicity, and in order to better compare with previous fits of eBOSS data, within this work we use a parameterization equivalent to the one assumed in \cite{Palanque+19} (except for the validation in \cref{ssec:l1o}): broken power laws with a pivot redshift $z_{\rm p}=3$ for the functions $T_0(z)$ and $\lambda_P(z)$, and power laws for the functions $\gamma(z)$ and $\tau_\mathrm{eff}(z) = - \ln \bar{F}(z)$. For instance, the temperature at mean density and the effective optical depth are parameterized explicitly as
\begin{equation}\label{eq:powerlaws}
T_0(z) = T_0(z_p) \left[\frac{1+z}{1+z_p}\right]^{{\eta_T}~(+\Delta \eta_T ~\mathrm{if}~z>z_p)} ~,
\quad
   \tau_\mathrm{eff}(z) = A_\tau \left[\frac{1+z}{1+z_p}\right]^{\eta_\tau}~.
\end{equation}
The cosmological parameters are of course not varied across redshift. For the parameters and priors related to this parameterization, see \cref{ssec:app:nuisance}.

\subsection{Likelihood and Inference procedure}
The likelihood is based on a simple Gaussian assumption with the covariance matrix taken from the original analysis performed in \cite{Palanque+19}. The computation of the theoretical flux power spectrum is performed using the emulator outlined in \cref{ssec:emu} with possible nuisance corrections (as for example required for \cref{ssec:nuisances}) added on top. The likelihood pipeline code is freely available in the \texttt{lym1d} package.

In order to perform the MCMC sampling we use the \texttt{MontePython v3.0} package \cite{Brinckmann+18}. The corner plots are produced with \texttt{liquidcosmo}, which is a free chain analysis and plotting utility available at \url{https://github.com/schoeneberg/liquidcosmo}.

\section{Emulator Validation}\label{sec:validation}

In this section we present the results of the validation of the emulator described in \cref{ssec:emu} applied to the \PoneD{} inference. We first perform a number of leave-one-out tests to gauge the inherent systematic uncertainty of the emulator and subsequently fit the \PoneD{} of the fiducial simulation to check if we correctly recover the cosmological and astrophysical parameters.

\subsection{Leave-one-out power spectra comparison}\label{ssec:l1o}

\begin{figure}
    \centering
    \includegraphics[trim={0cm 3cm 1cm 4cm},clip,width=0.7\textwidth]{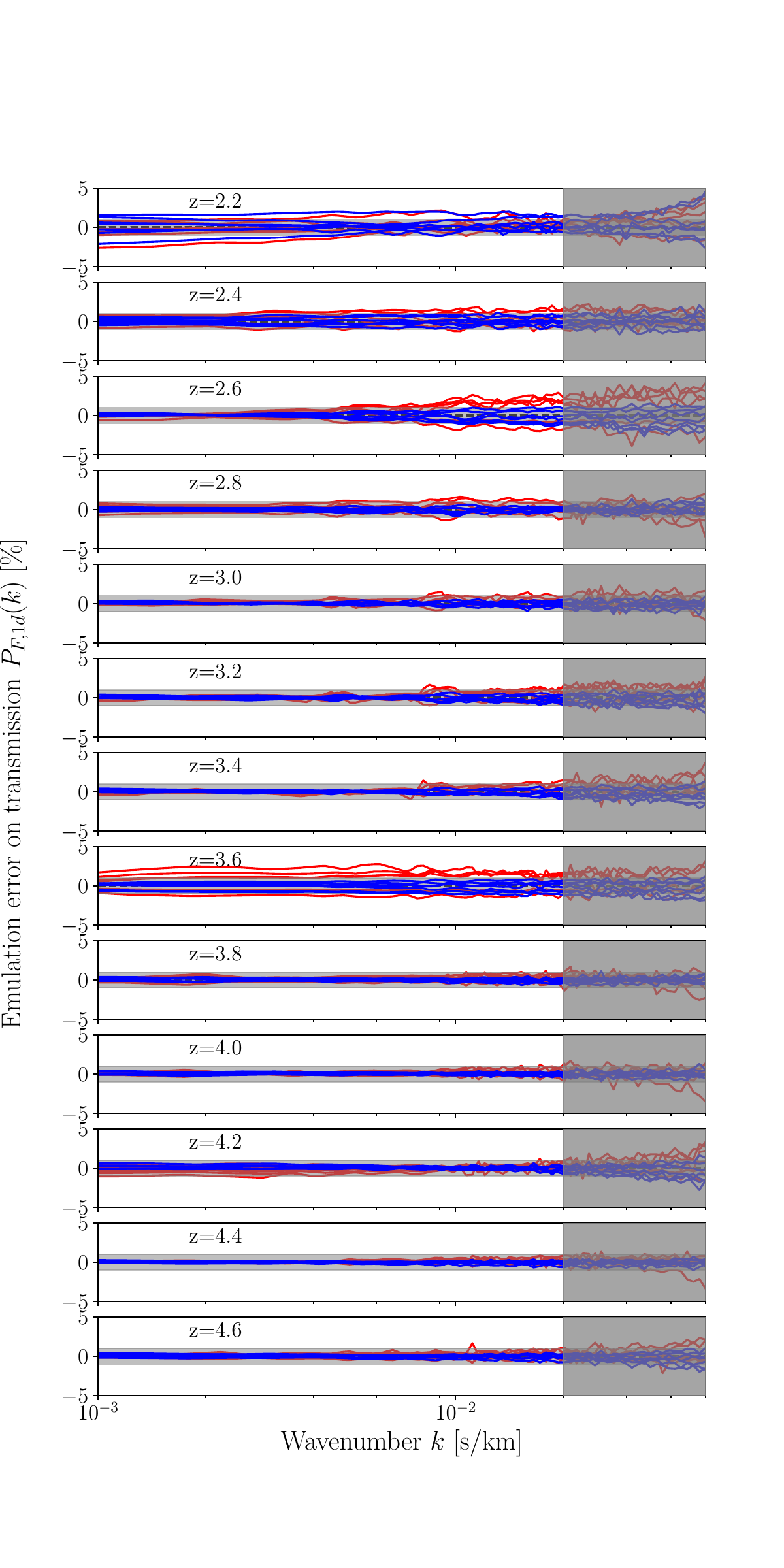}
    \caption{Results of leave-one-out tests. In each panel (corresponding to different redshift bins) we display the relative difference between the emulated one-dimensional flux power spectrum and that of the original simulation. The color of the line specifies whether the parameters of the left out model are on the edge of (red) or inside (blue) the cosmological parameter hypercube spanned by the other simulations (see text for explanations). The light grey horizontal bands show the $1\%$ error range. The dark grey band on the right is not accessible with eBOSS, but can be partially accessed with DESI.}
    \label{fig:leave-one-out power spectrum}
\end{figure}

In this first test we check that the emulator correctly reproduces the power spectrum of a given simulation when only the other remaining simulations are used to train it. Explicitly, all astrophysical post-processings of one simulation are removed from the training set while all post-processings of other simulations are kept. We then attempt to emulate the intrinsic astrophysical evolution of the left out simulation.
This allows us to estimate the accuracy of the emulator's predictions for the transmission power spectrum.

Note that this result actually quantifies the joint residual from two effects: a) the accuracy of the astrophysical post-processing when generating different realizations of \{$T_0$,\,$\gamma$,\,$\bar{F}$\} and b) the interpolation accuracy of the emulator in different parts of the cosmological parameter space\footnote{This also includes variations of the heating rate and thus emulation in $\lambda_P$\,.} in the worst-case scenario of reduced model density at the point of interest. 

\Cref{fig:leave-one-out power spectrum} shows the result of the leave-one-out test for redshifts between 2.2 and 4.6.
The accuracy of the emulation procedure is at the percent level on all relevant wavenumbers for the considered redshift bins as long as the left out simulation is not on the edge of the cosmological parameter space used for training. We quantify whether a left out simulation is \enquote{inside} or \enquote{on the edge} of the training volume by checking whether the smallest hypercube encompassing the parameters of all remaining simulations contains the parameters of the left out simulation or not.

At $z\leq2.6$, we find an emulation error of up to around $2\%$ for all models \enquote{inside} the training volume. The accuracy is better than $1\%$ (corresponding to the grey bands) at higher redshift. There is a notable exception at $z=3.6$, where a critical computing issue in one of the hydrodynamical simulations resulted in a corrupted snapshot at this redshift. This prevented the generation of the main suite of thermal models, such that only the auxiliary suite was available in this case. Attempts to re-simulate the missing snapshot are ongoing.

Our emulator is thus able to achieve good accuracy. We will now check whether this accuracy is sufficient for inferring cosmological parameters from mock data without significant bias. 

\subsection{Fitting the fiducial model}\label{ssec:fiducial}

\begin{figure}
    \centering
    \includegraphics[width=0.68\textwidth]{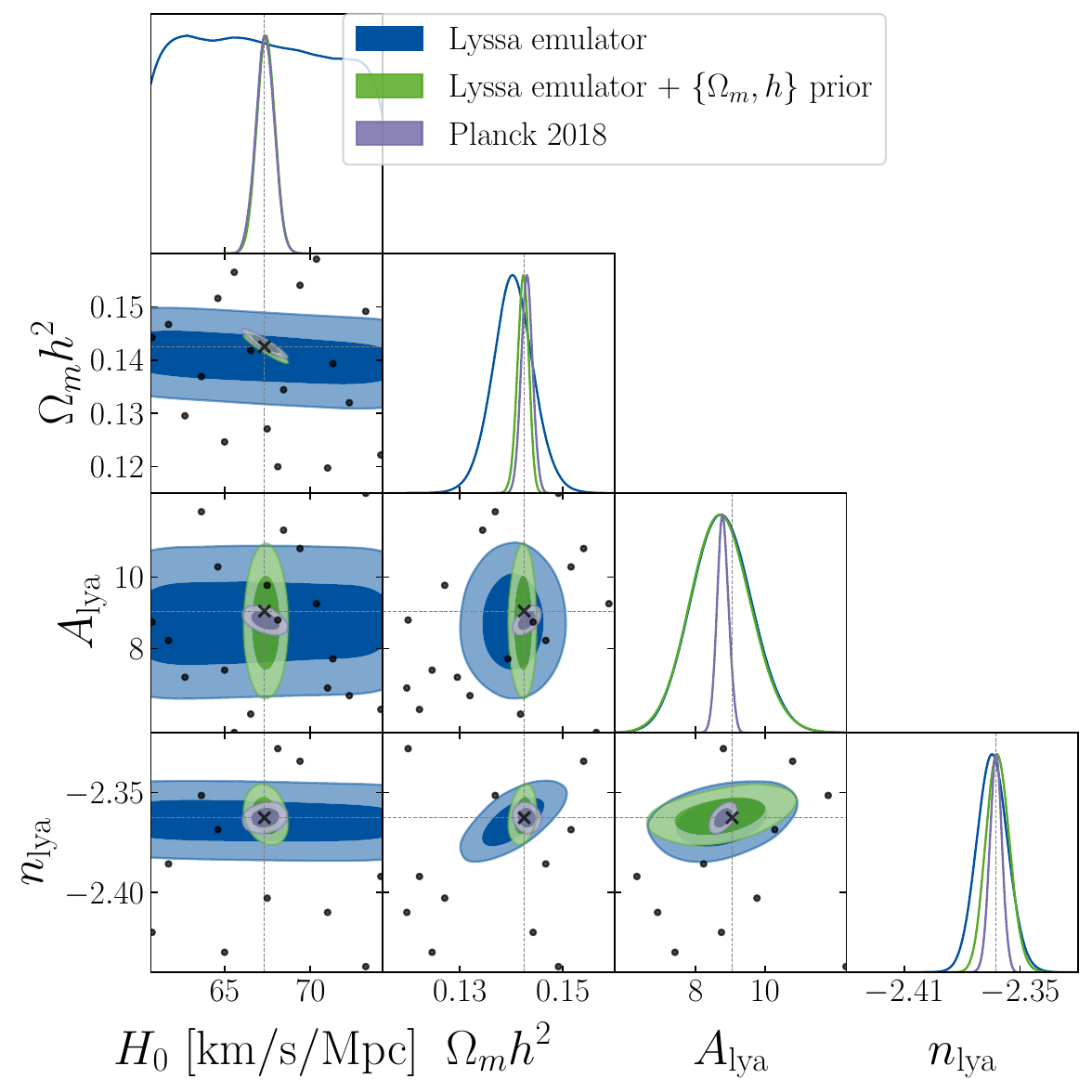}
    \caption{68\% and 95\% CL credible regions and marginalised posteriors when fitting mock data extracted from the fiducial simulation with the new \gridname{}-based emulator. In blue, we show the posteriors obtained with our baseline priors, summarized in \cref{tab:priors}. The green contours and posteriors include additionally a Gaussian prior on $\{\Omega_{\rm m}, h\}$ derived from Planck (skee text). The purple contours and posteriors, post-processed from the Planck chains, are shown for comparison. The black points represent the parameters of the simulation suite and the black crosses the parameters of the fiducial simulation (which are very close, but not exactly equal to the Planck mean values). The grey lines show the parameter values of the fiducial simulation. The fit is able to reconstruct these values without any significant bias. For correlations with nuisance parameters, see \cref{fig:fid_fit_nuisance}.
    }
    \label{fig:fid_fit}
\end{figure}
In this section, we check whether the emulator can recover the cosmological and thermal parameters in a full inference run, using some mock data created from a simulation that was not used to train the emulator. For this purpose, we use the fiducial simulation mentioned in \cref{ssec:design}. This test does not only check that the fiducial power spectrum is correctly recovered at fiducial parameter values, but also that there are no other parameter combinations in a different but nearby region of parameter space that predict the same power spectrum.

To create mock data, we measure the flux power spectrum in the fiducial simulation at the wavenumbers used in eBOSS measurements (in units of $\mathrm{s/km}$). We do not add astrophysical (DLA, metals, ...) or instrumental corrections to this mock data. In order to obtain some realistic uncertainties, we rely on the eBOSS data analysis. Thus, our mock likelihood incorporates the same covariance matrix as in \cite{Palanque+19}.

The results of this test are shown in \cref{fig:fid_fit}. The emulator recovers very well the cosmological parameters of the fiducial simulation, and in particular the amplitude $A_\lya$ and tilt $n_\lya$ to which \lya{} data are most sensitive. 
We even find some sensitivity to $\Omega_{\rm m} h^2$, while the posteriors on $H_0$ are almost flat, as expected given the arguments of \cite{Pedersen+19,Pedersen+23}. The thermal nuisance parameters are also nicely recovered, as shown in \cref{fig:fid_fit_nuisance}.

\begin{table}
    \centering
    \begin{threeparttable}[]
    \caption{Cosmological parameters of the simulations in our suite. The baryon density is fixed to $\Omega_b h^2 = 0.02233$. See also \cref{fig:parameter-grid}.}\label{tab:sim_parameters}
    \begin{tabularx}{\textwidth}{XcX}
    & \begin{tabular}{c|c c | r r | c}
        Simulation & $\Omega_{\rm m} h^2$ & $h$ & $A_\mathrm{\lya}$ & $n_\mathrm{\lya}$ & $A_\mathrm{UVB}$\\ \hline
        fiducial\tnote{*} & 0.1425 & 0.6732 & 9.0514 & -2.3626 & 1.0000 \\ \hline
        0 & 0.1516 & 0.6461 & 10.2857 & -2.3686 & 1.2200\\
        1 & 0.1589 & 0.7035 & 9.2571 & -2.2657 & 0.9036 \\
        2 & 0.1271 & 0.6748 & 9.7714 & -2.4029 & 1.3750 \\
        3 & 0.1344 & 0.6844 & 11.3143 & -2.3171 & 0.5893 \\
        4 & 0.1442 & 0.6078 & 8.7429 & -2.4200 & 1.5321 \\
        5 & 0.1369 & 0.6365 & 11.8286 & -2.3514 & 0.9821 \\
        6 & 0.1540 & 0.6939 & 10.8000 & -2.3343 & 1.6107 \\
        7 & 0.1418 & 0.6652 & 6.1714 & -2.4714 & 1.4535\\
        8 & 0.1320 & 0.7226 & 6.6857 & -2.3000 & 1.0607 \\
        9 & 0.1393 & 0.7131 & 7.7143 & -2.4543 & 0.7464 \\
        10 & 0.1565 & 0.6556 & 5.6571 & -2.2829 & 1.2964 \\
        11 & 0.1467 & 0.6173 & 8.2286 & -2.3857 & 0.6679 \\
        12 &  0.1295 & 0.6267 & 7.2000 & -2.2486 & 1.1393 \\
        13 & 0.1491 & 0.7322 & 12.3429 & -2.4371 & 0.8250 \\ \hline
        14 & 0.1197 & 0.7100 & 6.9000 & -2.4100 & 0.6600 \\
        15 & 0.1222 & 0.7400 & 6.3000 & -2.3920 & 1.5400 \\
        16 & 0.1246 & 0.7100 & 7.4000 & -2.4300 & 0.3200 \\
        17 & 0.1200 & 0.6810 & 8.8000 & -2.3280 & 1.0100 
    \end{tabular} & 
    \end{tabularx}
    \begin{tablenotes}
    \item[*] \footnotesize This simulation has been run using transfer functions from CAMB with $A_s=2.1\times 10^{-9}$ and $n_s=0.966$, but the corresponding values of $A_\mathrm{\lya}$ and $n_\mathrm{\lya}$ were derived from those parameters using CLASS. It also uses a slightly different box size of $80\,\hinvMpc\approx118.8\,\Mpc$ whereas the others use $120\,\Mpc$.
    \end{tablenotes}
    \end{threeparttable}
\end{table}

Since the emulator does not provide significant constraints on $H_0$ or $\Omega_{\rm m}$ individually, we also check whether setting priors in the $\{\Omega_{\rm m},h\}$ plane biases the results on $A_\lya$ or $n_\lya$\,. This check is performed in view of combining the Lyman-$\alpha$ data with data from CMB experiments such as Planck. We use a prior with a covariance matrix in $\{\Omega_{\rm m},h\}$ derived from the Planck 2018 TTTEEE+lensing results \cite{Planck:2018vyg}, centered at the fiducial simulation.

\Cref{fig:fid_fit} shows that with such priors we recover almost the same constraints, with improved precision by 20\% on $n_\lya$ due to its correlation with $\Omega_{\rm m} h^2$. We stress that such a correlation between $\Omega_{\rm m} h^2$ and $n_\lya$ is present in all of our results (see below), motivating a potential optimization of the parametrization in future studies.

\section{Comparison with pre-existing simulations}\label{sec:simulation-comp}

In this section, we compare the \gridname{} emulator to the existing B13\nobreakdash-based Taylor emulator and check how compatible they are within our likelihood framework. We start with introducing an additional caveat due to the approach we took in generating initial conditions, and develop a correction for this effect which has to be applied when comparing to external simulations or observational measurements. Afterward, we perform inference on the fiducial B13 model using both emulators.

\subsection{A caveat and its correction: Initial conditions}
\label{ssec:ic}
\begin{figure}
    \centering
    \includegraphics[width=0.50\textwidth]{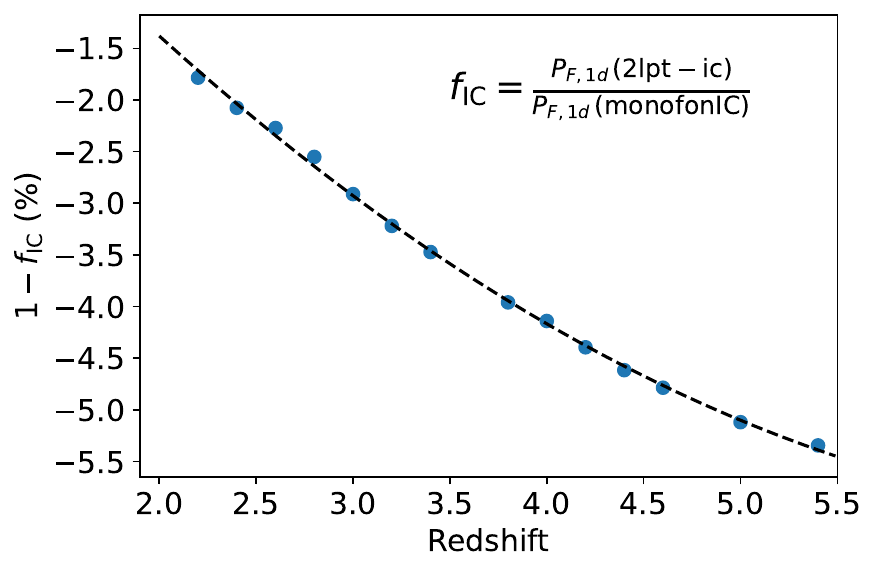}
    \includegraphics[width=0.47\textwidth]{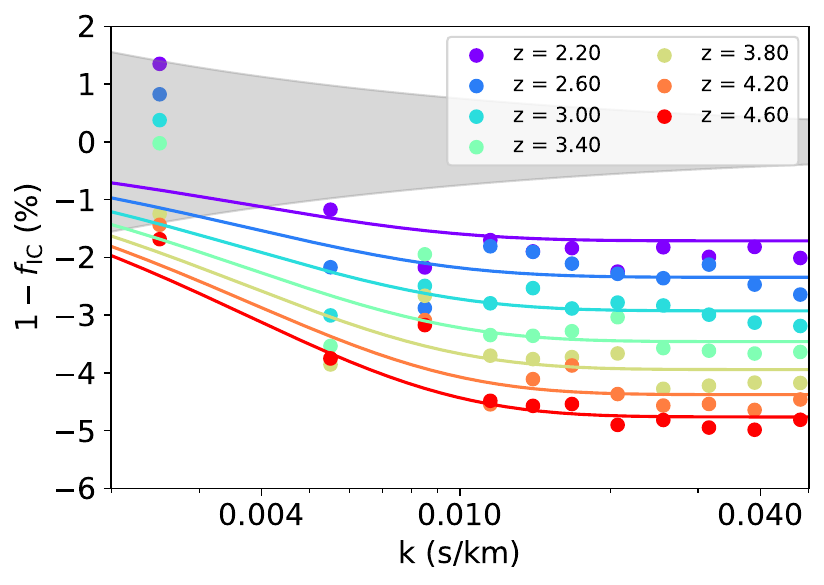}
    \caption{Difference in the \PoneD{} caused by using different initial conditions (one-particle transfer function vs split between baryons and dark matter). The corrections represent the relative difference in \% between the \texttt{monofonIC} run and the corresponding \texttt{2lpt-ic} simulation. Left: mean suppression as a function of redshift for $0.01<k<0.05$~s/km, where points are from simulations and the dashed line is the correction factor as parameterized by \cref{eq:fic}. Right: suppresion as a function of $k$, plotted for a subset of redshifts, with the continuous lines from \cref{eq:fic}. The shaded band represents an estimation of the $1\sigma$ cosmic variance associated to the finite box size. 
    }
    \label{fig:IC}
\end{figure}

As described in \cref{ssec:individual_sim}, initial conditions were generated with the \texttt{2lpt-ic} code, using a single particle species with the total matter transfer function at $z_{\rm ini}=99$, instead of two species, with the transfer functions of baryons and dark matter respectively.\footnote{These differ due to non-adiabatic perturbations seeded by the pre-recombination acoustic oscillations \cite{Schmidt:2016}. The species differ in both their over- and underdensities and velocities.} To quantify the impact of the choice of initial condition code and setting, we ran an additional simulation using the \texttt{MUSIC2-monofonIC} code \citep{Hahn+20,Hahn+2020_software}, generating two-fluid initial conditions, while using the same seed and random number generator as our \texttt{2lpt-ic} runs.

We find that, after rescaling mean optical depths to a common value, the predicted flux power spectra differ by a few percent, as shown in \cref{fig:IC}. The main difference is an overall suppression of the power, with a smooth evolution as a function of redshift: this suppression reaches 5~\% at high redshift, but it is 2~\% for $z\simeq 2.4$, where the precision of \PoneD{} measurements is highest. As expected, the impact of separating out baryons and dark matter initial conditions is larger at high redshift and reduces at later times. We checked that this suppression does not vary significantly when post-processing the simulations to emulate different thermal histories. The suppression does not vary as a function of the wavenumber $k$ for $0.01 < k < 0.05$~s/km. The results indicate that the amplitude of the suppression is modified for smaller values of $k$, as one may expect, but the exact dependence as a function of wavenumber on large scales is difficult to quantify due to the finite box size effect, and simulation-related noise that manifests as oscillations of the flux power spectrum.
 
We correct for this effect using an analytical correction as a multiplicative factor $f_\mathrm{IC}$ to the emulator's predicted \PoneD{}:
\begin{equation}
    f_\mathrm{IC}=1 - 0.01\times[ Az^2 + Bz + C] [1 - \exp(-k/k_\mathrm{IC})]
    \label{eq:fic}
\end{equation}
with $k_\mathrm{IC} = 0.0037\,\mathrm{s/km}$, $A=0.15$, $B=-2.3$, $C=2.6$. The main outcome of this correction is a redshift-dependent offset relevant at all wavenumbers as well as a subdominant exponential part that is relevant for $k \lesssim 0.01 \mathrm{s/km}$. Note that the modeled range in $k$ is sufficient for correcting all measured scales in BOSS ($k<0.02\,\skm$) and DESI ($k\lesssim 0.05\,\skm$).

We do not attempt to model the noise-related oscillations. We enable the correction $f_\mathrm{IC}$ when comparing the emulator with \Borde simulations in \cref{ssec:previous}, and in the application to data shown in \cref{sec:data}. We also explicitly test the impact of not using the correction (see \cref{fig:data_fit}, right panel). In \cite{Fernandez:2020jgf} the authors investigate the impact of using a single-species or two-species initial conditions on the \texttt{gadget}-based Lyman-$\alpha$ forest. A more detailed investigation into the effect of initial conditions generated from different codes on the flux power spectrum in the context of \texttt{Nyx} is left for future work.

\subsection{Fitting previous \texorpdfstring{\Borde}{B13} simulations}\label{ssec:previous}

We additionally wish to test whether our new emulator is able to provide a reasonable fit to the power spectra extracted from one of the \Borde simulations presented in \cite{Borde+13,Palanque+19}. For this test, we choose the central simulation that was used as the expansion point of the old Taylor emulator. Our purpose is to assess whether the improved modeling of the flux power spectrum presented in this work is likely to change the cosmological parameter constraints derived from previous analyses.

In principle, there could be many reasons for which the new emulator would fail to extract the correct parameters from the old \Borde simulation.
First and foremost, the new simulations are run with much higher dynamic range, using $4096^3$ dark matter particles and hydrodynamical cells each, whereas in \cite{Borde+13} simulations were spliced together from three simulations, two with $2\times768^3$ particles with a large and a small box, and one with $2\times192^3$ particles in a small box at the same resolution as the large one. This approach required a splicing correction which was marginalized over in previous analyses, and is not needed anymore when using a larger dynamic range simulation as in the \gridname{} suite. Additionally, fundamentally different codes were used for the simulations (SPH-based \gadget compared to grid-based \texttt{Nyx}). The difference between these simulations is well described in \cite{Walther+20}, showing that there is a growing deviation in the flux power spectrum towards smaller scales. Across the eBOSS scales adopted for this test ($k < 0.02\mathrm{s/km}$), this deviation can be approximated as a constant offset plus a difference in slope for higher redshifts. Thus, we may find a small reconstruction bias for the amplitude parameter $A_\lya$ and/or the mean transmission, $\bar{F}(z)$.

We also check here whether adding a Planck prior on $\{\Omega_{\rm m}, h\}$ -- centered on the \Borde central values -- leads to better constraints or biases the results. This is to estimate if the cosmological constraints would be significantly changed by combining with CMB data.

\begin{figure}
    \centering
    \includegraphics[width=0.6\textwidth]{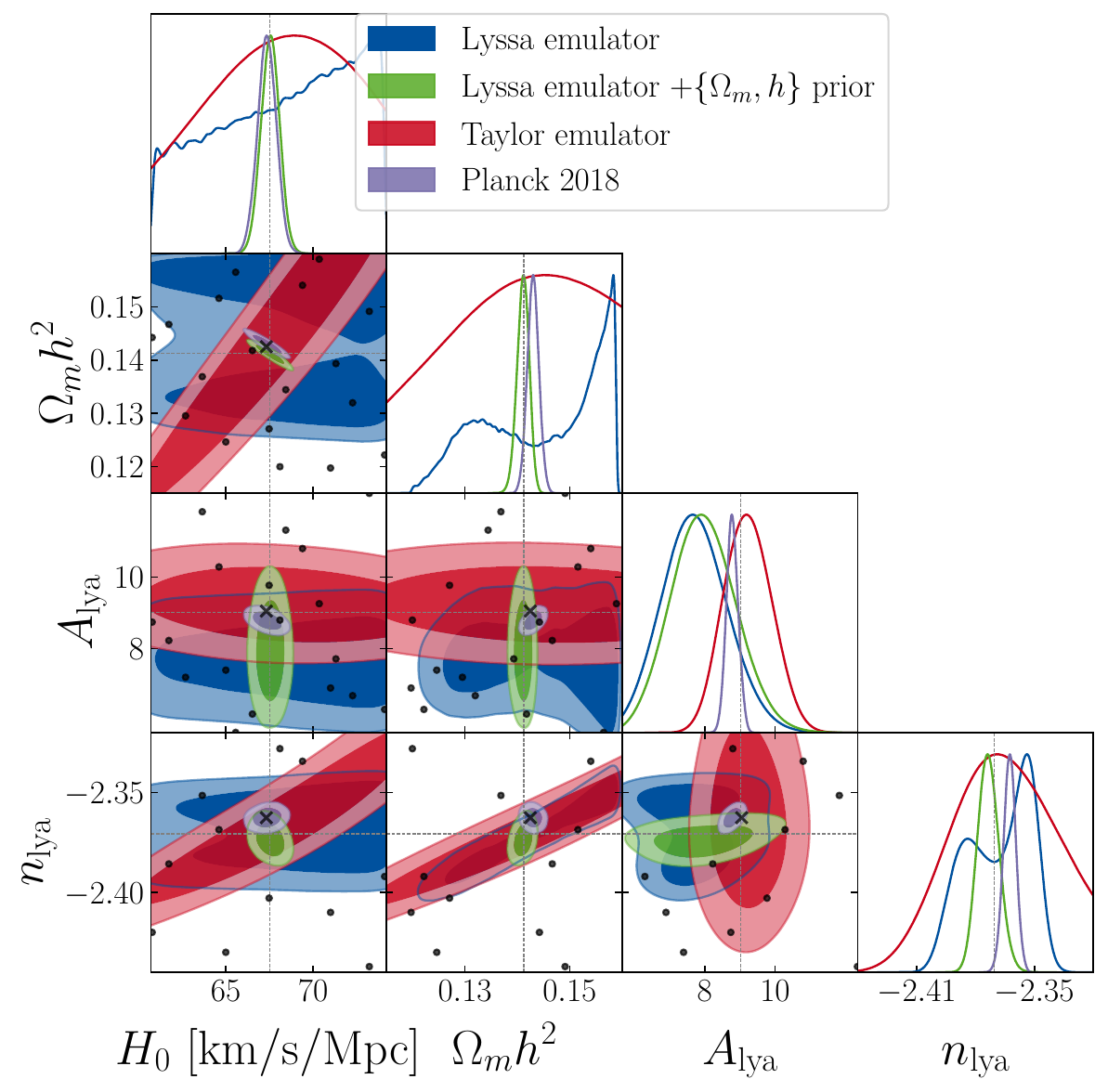}
    \caption{Same as in \cref{fig:fid_fit}, but fitting instead the old \Borde{} central simulation, whose parameters are represented by the grey lines. The fits with the \gridname{} emulator are shown in blue and green. We also include results using the Taylor emulator of \cite{Palanque+19} as a reference in red.}
    \label{fig:gadget_fit}
\end{figure}

Finally, we compare the contours obtained with our new \gridname{}-based emulator with those from the old Taylor emulator for this mock data, post-processed into the new parameter space.
\Cref{fig:gadget_fit} shows that, despite different codes and models of the power spectrum, the constraints obtained with the \gridname{}-based emulator are broadly consistent with the fiducial values assumed in the \Borde{} simulation.

There are a few notable differences compared to before: 
\begin{enumerate}
    \item The constraint obtained from the mock data with the \gridname{}-based emulator is now somewhat degraded compared to the results of the previous section. The largest change is that the posterior on $\Omega_{\rm m} h^2$ is flat and wide, with a very mild bimodality and an extremely mild preference for large values of $\Omega_{\rm m} h^2$ towards the edge of the design. This also leads to a degraded sensitivity on $n_\lya$ due to their correlation. The constraints orthogonal to the $\Omega_m h^2-n_\lya$ correlation direction or for $A_\lya$ do not really degrade.
    \item The constraints on the $A_\lya$ and $n_\lya$ parameters are consistently recovered with a mild $-1.4\sigma$ downward bias on $A_\lya$ ($n_\lya$ is recovered with only a $+0.15\sigma$ bias).
    \item Since the prior on $\{\Omega_{\rm m}, h\}$ breaks the $\Omega_{\rm m} h^2 - n_\lya$ degeneracy, the $n_\lya$ constraints become significantly tighter once the prior is added. The slope remains almost unbiased with the prior ($-0.3\sigma$ bias), while the bias of the amplitude is reduced ($-0.9\sigma$ bias). 
    \item The degeneracy between the $n_\lya$-$h$ parameters as well as the $\Omega_{\rm m} h^2$-$h$ parameters is somewhat different in the Taylor emulator case, which we attribute to the finite accuracy of the employed method of using single cross-derivatives of the Taylor emulator and to the fact that the Taylor expansion was limited to second order, albeit a more detailed investigation is left for future work.
\end{enumerate}

\enlargethispage*{2\baselineskip}
Overall, we notice that despite the very different codes the cosmology is recovered very consistently (apart from a slight $\sim 1\sigma$ bias in the amplitude). For this reason, we do not expect a strong difference between the constraints on cosmological parameters obtained with the Taylor and \gridname{}-based emulator when applied to real data. We also stress that this $\sim 1\sigma$ bias on $A_\lya$ is reduced  when adding a prior on $\{\Omega_{\rm m}, h\}$ from Planck.

\section{Fitting observational data}\label{sec:data}
In this section we aim to fit the observational data for \PoneD{} of SDSS eBOSS DR14 presented in \cite{Palanque+19}. The goal is both to compare the inferred parameter values between the old Taylor emulator and the new \gridname{}-based emulator, as well as to investigate a variety of settings and analysis choices with regard to the impact of such choices on a real data analysis. We first discuss our treatment of nuisance parameters, summarize a baseline analysis including tests of using different parametrization and corrections, and finally show how additional informative priors (on flux and/or cosmological parameters) change the given results.

\subsection{Additional analytical corrections for observational nuisances}\label{ssec:nuisances}

In order to fit eBOSS data, we have to include a number of nuisance parameters meant to marginalize over astrophysical and instrumental effects.
These corrections are purposefully kept identical to those of \cite{Palanque+19} to ease the comparison, while ones more suited for DESI will be developed in future work. We also apply the correction discussed in \cref{ssec:ic}. The final power spectrum incorporates all corrections as
\begin{equation}
    \text{\PoneD} =  f_\mathrm{metals} \,\frac{\left(P_\mathrm{F,emu}+f_\mathrm{noise} P_\mathrm{noise}\right)}{f_\mathrm{DLA} \, f_\mathrm{SN} \, f_\mathrm{AGN} \, f_\mathrm{IC}}
\end{equation}
with all quantities depending on both $k$ and $z$. We use the following corrections:
\begin{enumerate}
    \item The \PoneD~measurement relies on the subtraction of a noise component $P_\mathrm{noise}$ inferred from the spectrograph noise. We add a rescaled (by a factor $f_\mathrm{noise}$) version of this component, accounting for incomplete or overzealous subtraction. The wavenumber independent prefactor is left free for each redshift bin independently, and marginalized over. 
    \item 
    The simulations do not model the presence of Damped Lyman-$\alpha$ Absorbers (DLAs), that are associated to the circumgalactic medium of (proto)-galaxies close to the line-of sight. While a fraction of those DLAs could be removed from the SDSS data used to estimate \PoneD, there is a residual contamination. Here we use a correction factor (originally based on \cite{McDonald+05b} and \cite{Viel+04})
    \begin{equation}
        f_\mathrm{DLA} = 1 - \left(\frac{1}{15\,000 \, k -8.9} +0.018\right) \, 0.2 \, A_\mathrm{DLA}~, \label{eq:dla}
    \end{equation}
    where we marginalize over a single parameter $A_\mathrm{DLA}$. 
    \item Although most of the signal in the Lyman-$\alpha$ forest originates from the pristine IGM, feedback processes from galaxies (such as supernovae, AGN winds, etc.) also contaminate the flux power spectrum. We parameterize this effect with the updated correction factor
    \begin{equation}
        f_\mathrm{SN} = 1 + A_\mathrm{SN} \, \left[d_0 + (d_1-d_0) \, \frac{k-k_0}{k_1-k_0}\right]~,\label{eq:sn}
    \end{equation}
    with $k_0=0.001 \,\mathrm{s/km}$, $k_1 = 0.02 \,\mathrm{s/km}$, $d_0 \in \{-0.06,-0.04,-0.02\}$ (for $z<2.5$, $2.5<z<3.5$ and $z>3.5$), $d_1 = -0.01$, and we marginalize over a single parameter $A_\mathrm{SN}$ \cite{2013MNRAS.429.1734V}.
    \item We define an analytical correction for the presence of AGN (also discussed in \cite[][Table 4]{Chabanier+2002.02822}) as
    \begin{equation}
        f_\mathrm{AGN} = (1+d(k,z) \, A_\mathrm{AGN})^{-1}~,\label{eq:agn}
    \end{equation}
    where $d(k,z)$ is a polynomial in redshift with a $k$-dependence of the form $a+b \exp(-c \, k)$. The parameters $(a,b,c)$ were previously fitted to full Galaxy formation simulations including AGN and SN feedback prescriptions from the Horizon-AGN and Horizon-noAGN runs \cite{Dubois+14:1402.1165}. They have been determined independently at each redshift and are kept fixed in our analysis while a single parameter $A_\mathrm{AGN}$ is marginalized over.
    \item We marginalize over a correction for silicon absorption lines,
    \begin{equation}
        f_\mathrm{metals} = \prod_i \left(1+a_i^2 + 2 a_i \cos(k \Delta v_i)\right)~,\label{eq:metal}
    \end{equation}
where $\Delta v_\mathrm{\SiIII}=2271\, \mathrm{km/s}$, $\Delta v_\mathrm{\SiII}=5577\,\mathrm{km/s}$, the amplitude of each transition is given by $a_i=f_i/(1-\bar{F})$,\footnote{In order to ease comparison with the previous approach of \cite{Palanque+19}, in this particular formula, we fix the evolution of $\bar{F}(z)$ to the same \textit{fiducial} evolution as in this reference, instead of matching it to the mean transmission function of the given model.}%
    and a single parameter $f_i$ is marginalized over (one for SiIII, and one for SiII).
    Note that this correction is needed only for absorption lines with restframe wavelengths comparable to the Lyman\nobreakdash-$\alpha$ transition as the respective metal forest then overlaps the probed Lyman\nobreakdash-$\alpha$ forest leading to correlations in the absorption field. For metal transitions with restframe wavelengths $\lambda_\mathrm{rest}>1270 \, \text{\AA}$ an estimation and subtraction of the metal power have been performed instead at the measurement level, specifically in the observed \PoneD, see \cite{McDonald+06,Palanque+13,Chabanier+19}.
\end{enumerate}
We expect many of these corrections to be insufficiently accurate for the analysis of upcoming survey data. However, they are well suited for performing an apple-to-apple comparison between the eBOSS fits performed with our new emulator and with the old Taylor emulator of~\cite{Palanque+19} as well as the results of~\cite{Rogers+23}. Future works will adopt more extensive corrections. In \cref{tab:priors} and \cref{ssec:app:nuisance}, we summarize the set of nuisance parameters used in our analysis and the exact priors assumed on these parameters. These priors are nearly equivalent to those assumed in reference \cite[Tab.~2]{Palanque+19}.

The Taylor emulator of~\cite{Palanque+19} includes additional corrections for splicing and background UV fluctuations. In the new emulator, the former are irrelevant since we no longer use splicing, and the latter we neglect -- in the context of the present work -- since the fit of ~\cite{Palanque+19} was compatible with no such fluctuations.

\subsection{Application to eBOSS data -- baseline analysis}\label{ssec:eBOSS_data}

\begin{figure}
    \centering
    \includegraphics[width=0.65\textwidth]{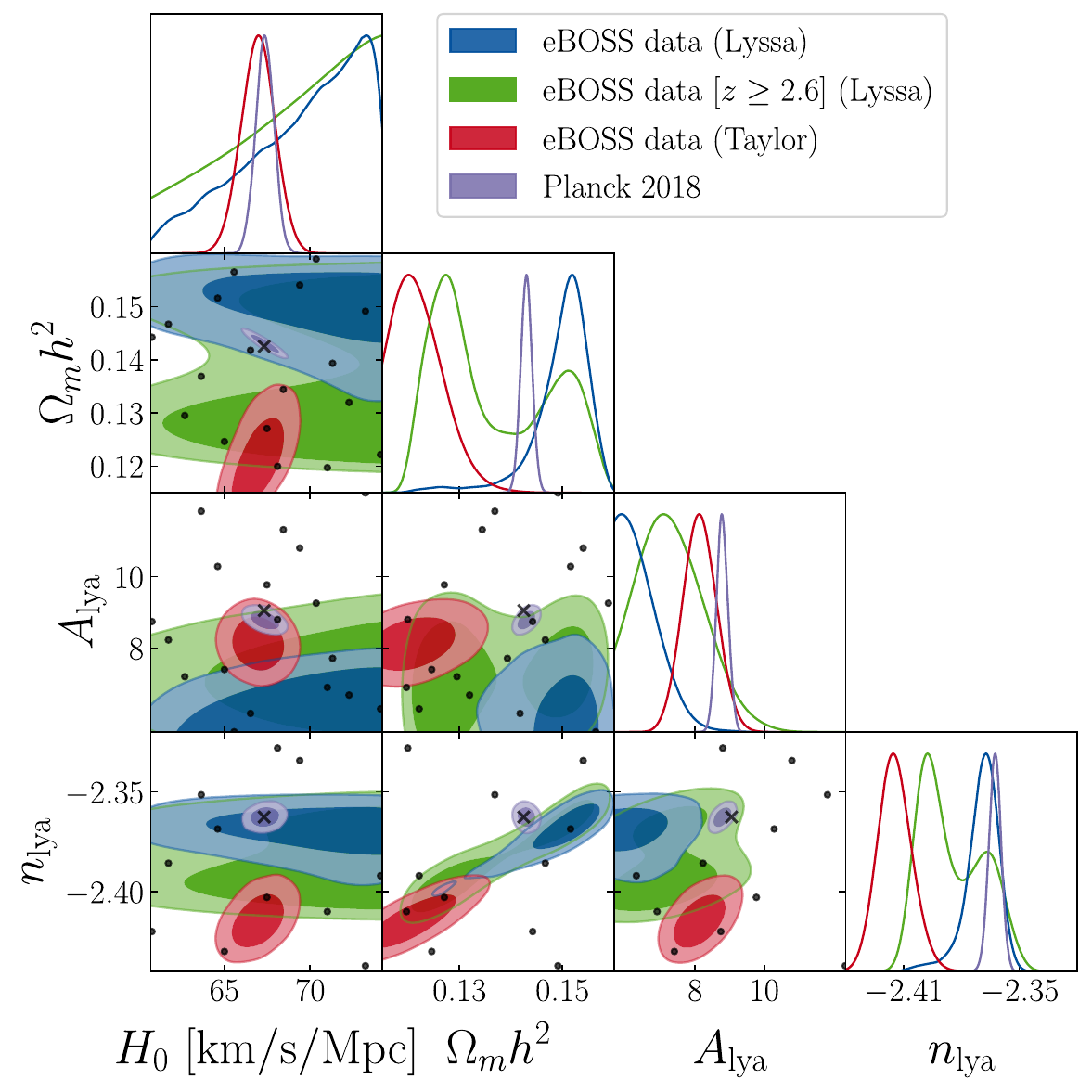}
    \caption{68\% and 95\% confidence contours and marginalized posteriors on cosmological parameters, obtained by fitting eBOSS data with the \gridname{}-based emulator. In blue we show the nominal run, in green a run with only data at $z\geq 2.6$, in red a run using the Taylor emulator of \cite{Palanque+19} (with an additional prior on $H_0$ as was used in the original chains), and in violet the constraints re-processed from Planck data assuming a flat $\Lambda$CDM cosmology. Note that, as we show in the right panel of \cref{fig:data_fit_FP}, the results in $A_\mathrm{lya}$ and $n_\mathrm{lya}$ do not strongly change when a prior in $H_0$ is considered for the \gridname{}-based emulator. The contours and posteriors involving nuisance parameters can be found in \cref{fig:data_fit_nuisance,fig:data_fit_nuisance_2}.}
    \label{fig:data_fit} 
\end{figure}

We now use our \gridname{}-based emulator to fit the current eBOSS data, which are described in~\cite{Chabanier+19,Palanque+19}. In this case, all the nuisance parameters described in \cref{ssec:nuisances} and summarized in \cref{tab:priors} are marginalized over. For the Taylor emulator we also include a prior on \mbox{$H_0=(67.3\pm1)\mathrm{km/s/Mpc}$} to make the results equivalent to those of \cite{Palanque+19} for the massless neutrino case.\footnote{We mention that for the Taylor emulator we have checked that we recover the constraints of \cite{Palanque+19} in the original parameter basis (and have indeed checked that the likelihoods agree at the $<0.1\%$ level at a number of points). That $A_\lya$ in the constraints is lower than that of Planck 2018, while in \cite{Palanque+19} $\sigma_8$ was/is larger comes down to the fact that the scales at which the amplitudes are measured are very different. With very low $n_s$/$\Omega_m$ values (as in \cite{Palanque+19}), even a large $\sigma_8$ leads to a low $A_\lya$ at the corresponding scale. Similarly, while in \cite{Palanque+19} the discrepancy in $n_s$ is only at the 1-2$\sigma$ level, for $n_\lya$ it is clearly larger, again due to the different scale and the low $\Omega_m$ value found in \cite{Palanque+19}. These effects have also been pointed out in \cite{Rogers+23}. Explicitly, for $\sigma_8=0.82$, $n_s=0.96$, $\Omega_m=0.27$ (corresponding roughly to the mean of \cite{Palanque+19}), we find $A_\lya=8.2$ and  $n_\lya=-2.4$, in good agreement with \cref{fig:data_fit}.}

The results of the nominal analysis in which we use the emulator as described above are displayed in \cref{fig:data_fit}. These nominal results suggest that the data prefer a value of $A_\lya$ significantly lower (by $-3.8\sigma$) than the value extrapolated from Planck 2018 in the $\Lambda$CDM model. They also point at
a consistent though somewhat higher value of $\Omega_{\rm m} h^2$ (higher by $1.7\sigma$) and at a value of $n_\lya$ fully consistent with Planck results \cite{Planck:2018vyg} (lower by $-0.7\sigma$). Overall, this nominal analysis appears to point to a tension between the Planck CMB anisotropy data
and the eBOSS Lyman-$\alpha$ data when a $\Lambda$CDM model is assumed. This is reminiscent of what has been found already in \cite{Palanque+19} and \cite{Rogers+23}. We notice that one major difference with respect to these previous analyses is that the tension appears in $A_\lya$, and not in $n_\lya$ (see \cref{fig:data_fit}, blue contour compared to red contour, compare also \cite[Fig.~1]{Rogers+23}). We also note that, somewhat similar to \cref{ssec:previous}, $\Omega_{\rm m} h^2$ is poorly constrained by \lya{} data alone (as expected from \cite{Fernandez+23}). 
Indeed, in the constraints reported here and in \cref{ssec:results_priors}, we observe that the data do not exhibit a strong $\Omega_{\rm m} h^2$ feature, leading to only very degraded or even bimodal constraints in $\Omega_{\rm m} h^2$. For this reason we also check in \cref{ssec:results_priors} if a prior on $\Omega_{\rm m} h^2$ significantly impacts these conclusions.

To investigate the tension in $A_\lya$ further, we check whether it could be induced by the well motivated choices in the construction of our emulator, for example the assumption of a broken powerlaw redshift dependence of $\lambda_P$\,, see \cref{app:ssec:free_thermal}. In none of the test cases we find a relaxation of the tension in $A_\mathrm{lya}$\,.
As such, it may seem that a tension similar to that reported in \cite{Palanque+19,Rogers+23} is confirmed. However, as we show below, there are several important caveats to such a conclusion.

\enlargethispage*{2\baselineskip}
As a first indication, we note that the removal of data below redshift $z<2.6$ as advocated in~\cite{Fernandez+23} reduces the tension in $A_\mathrm{lya}$ significantly to $1.5\sigma$ (largely due to larger uncertainties and a slight shift in mean value by around +1$\sigma$). As such, the tension is disproportionately strongly driven by the lowest 2 out of 13 redshift bins, casting its physical nature into doubt. We do not conclude from our fits that there is a problem with the data at $z < 2.6$, as we discuss and show in \cref{ssec:results_priors}.

By investigating the correlations with nuisance parameters (see \cref{fig:data_fit_nuisance}) we managed to identify that there is a strong correlation between the amplitude $A_\lya$ and the overall amplitude of the observed mean transmission, where in particular the $A_\tau$ parameter is highly degenerate with $A_\lya$ and appears to prefer values close to the lower boundary of the prior. This result puts the accuracy of emulation into question in this regime (as we have seen in \cref{ssec:l1o}, the emulation accuracy degrades towards the edges of the design). As such, it is essential to check how the constraints change when imposing a well motivated prior on the mean transmission.

\subsection{Variations with more informative priors}\label{ssec:results_priors}

We have seen above that the critical degeneracies for cosmological analyses are the one between $A_\lya-A_\tau$ and between $n_\lya-\Omega_m h^2$\,. Luckily, for both of these degeneracies there are independent measurements that can be exploited. For $A_\tau$ there are independent measurements of the mean transmission, which we adopt in \cref{ssec:mean_trans}. Instead, for $\Omega_m h^2$ we can add other cosmological information, such as from Planck CMB observations, as we do in \cref{ssec:planck_prior}.

\subsubsection{Mean transmission prior}\label{ssec:mean_trans}
\begin{figure}
    \centering
    \includegraphics[width=0.47\textwidth]{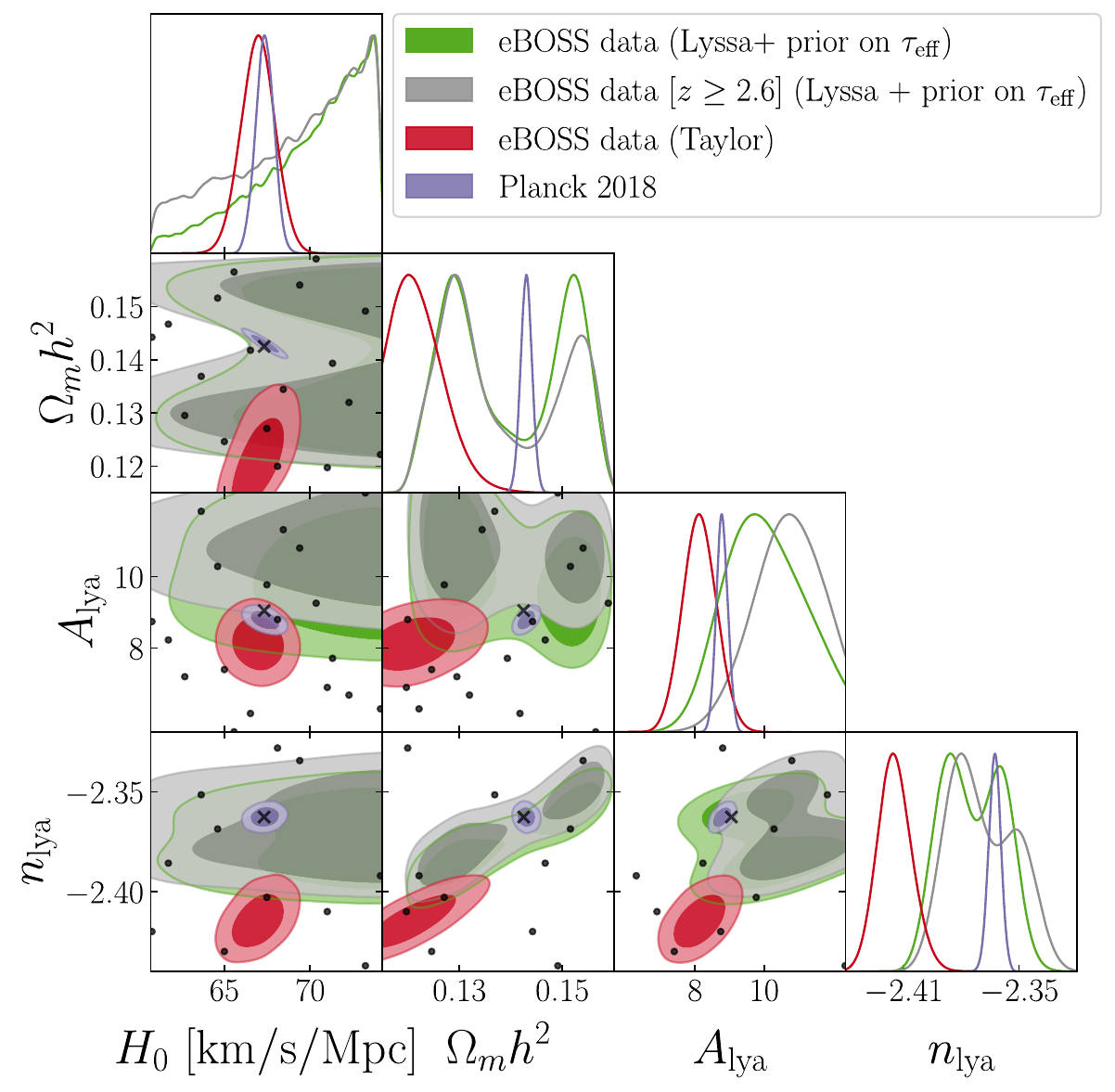}
    \includegraphics[width=0.51\textwidth]{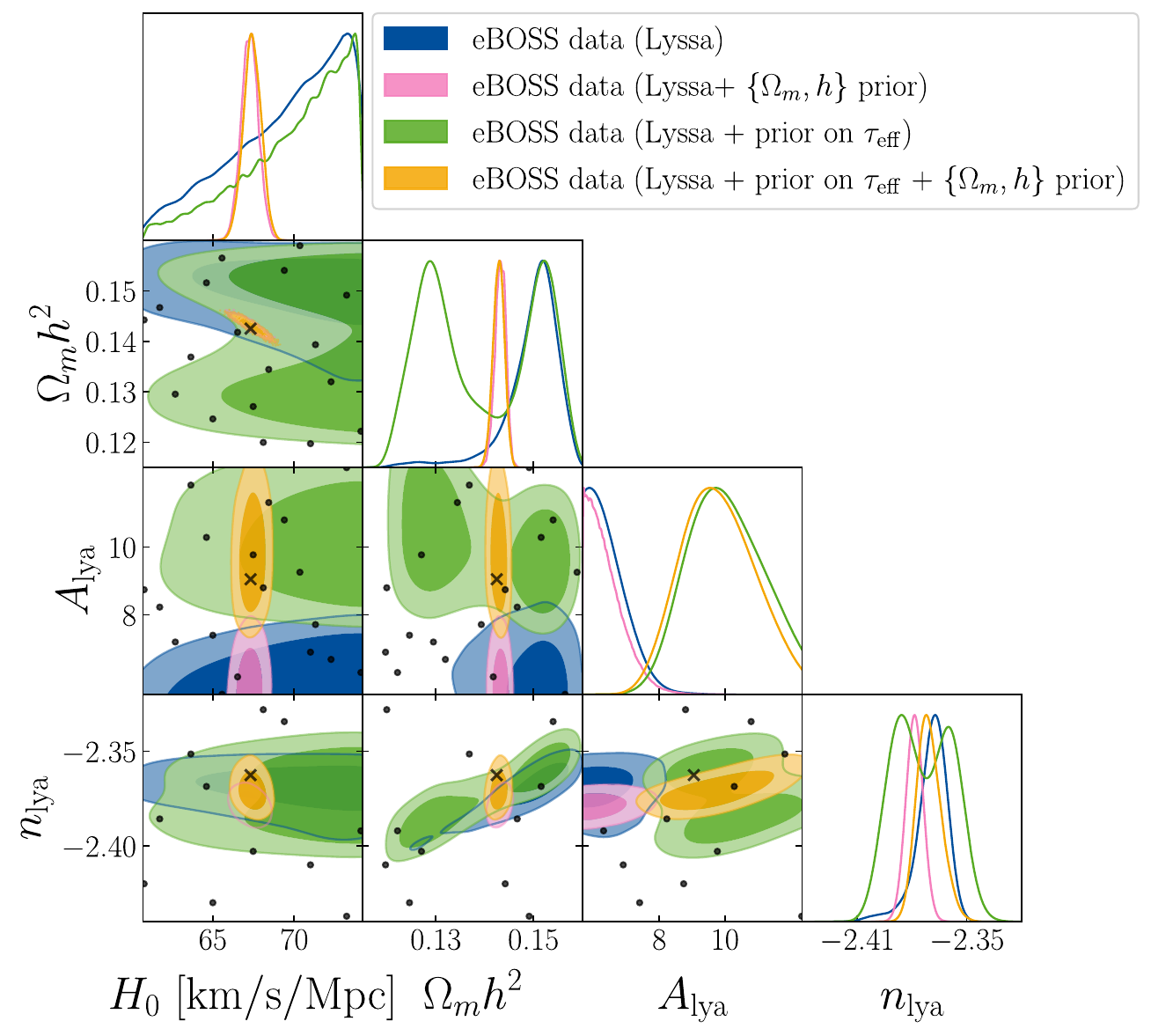}
    \caption{Same as \cref{fig:data_fit}, but including additional priors. Left: With an additional prior on the mean transmission from \cite{Becker+13}. Right: With an additional prior on $\{\Omega_{\rm m}, h\}$ from Planck, with or without the same additional mean transmission prior. Nuisance parameters can be found in \cref{fig:data_fit_FP_nuisance,fig:data_fit_FP_nuisance_2}. Note that as in \cref{fig:data_fit} the red contours include a prior on $H_0$\,.}
    \label{fig:data_fit_FP}
\end{figure}

To force the fit to remain in a regime that is well emulated, we impose a Gaussian prior on the observed effective optical depth $\tau_\mathrm{eff}(z)$ (which is related to the mean transmission by $\bar{F}(z) = \exp[-\tau_\mathrm{eff}(z)]$) according to the largely model independent results from \cite{Becker+13} at each individual redshift. While this measurement was performed from a sample of $\sim 6\,000$ SDSS spectra which overlaps with that used in~\cite{Chabanier+19}, it relies essentially on an anchor point from high-resolution spectra~\cite{faucher-giguere+2008}, and on a method to estimate quasar continua fundamentally different from that used in~\cite{Chabanier+19}. Therefore, within the scope of this study we can consider that the measurement from~\cite{Becker+13} is largely independent of the eBOSS \PoneD{}. 
Note that it is well known that the mean transmission strongly affects \PoneD{} (and is in particular degenerate with $A_\mathrm{lya}$). The idea to use external $\tau_\mathrm{eff}$ measurements in order to ease the determination of the amplitude of matter power spectrum from \PoneD{} is not new at all, see for example \cite{Seljak03}, although several recent analysis like~\cite{Palanque+19, Fernandez+23} do not use such an external prior. 

\Cref{fig:data_fit_FP} (left panel) displays our result when using this additional prior based on \cite{Becker+13}. The tension in $A_\mathrm{lya}$ completely disappears in this case ($+0.3\sigma$) with all data included. However, subsets of the data still can show a tension (such as $+1.2\sigma$ when only including data at $z\geq2.6$). We therefore conclude that, if the emulator is forced to remain within a regime of a mean transmission that agrees with independent measurements from \cite{Becker+13} and thus lies away from the edge of its prior range, then the tension in $A_\lya$ completely disappears.

We also note that this same shift (in the $A_\mathrm{lya}$-$A_\tau$ degeneracy direction) is what is observed in \cite[Tab.~3]{Fernandez+23} when cutting the data to $z\geq2.6$. We argue that at least in the case of the Lyssa emulator applying well-motivated priors on the mean flux can restore concordance of the cosmological constraints between high- and low-redshift data. We note, however, that unlike \cite{Fernandez+23} we do not see a big difference in the flux slope parameter or the primordial slope when fitting all redshift bins.

\enlargethispage*{1\baselineskip}
Given the differences between the analyses with and without a prior on the observed mean flux, we caution against strong interpretations of the baseline $3.8\sigma$ tension in $A_\lya$/$n_\lya$ (such as in~\cite{Rogers+23,Goldstein+23}), and instead argue to use the results when including a prior on the mean transmission from \cite{Becker+13}, until further investigations are performed.

A comprehensive comparison to \cite{Bird+23} and an investigation into the degeneracy of the observed mean flux with the cosmological parameters and corresponding estimates of systematic uncertainties is left for future work. We also note that other mean transmission priors (such as based on the recent measurement by \cite{TurnerWynne+24}) would lead to slightly different constraints on $A_\lya$ as well.

We also note that from \cref{fig:data_fit_FP_nuisance} that we see a large anti-correlation between $A_\lya$ and $T_0$, as well as a strong correlation between $T_0$ and its slope parameters $\eta_T$ and $\Delta \eta_T$. 
In principle also priors on $T_0(z)$, $\gamma(z)$ or $\bar{F}(z)$ based on high-resolution \lya{} forest observations such as \cite{Walther+19,Gaikwad:2020eip} could be applied\finalr{, but note that}{}\final{. However, note that} similarities \final{in the modeling} between the approaches taken in these measurements and here\final{, e.g. concerning the approaches towards simulation post-processing,} could lead to correlated methodological systematics\finalr{, that}{. Also,} parts of \final{the BOSS} dataset used in \cite{Walther+19} \final{alongside higher resolution data} overlapped with the eBOSS data studied in here, and \finalr{that}{} these analyses were performed \finalr{at fixed cosmological parameters}{ using smaller $10-20\,\hinvMpc$ simulation boxes and fixed initial conditions, i.e., fixed cosmological parameters and cosmic variance}. Although such priors could be used for strengthening the constraints and reducing prior volume effects on our cosmological measurement, to be conservative, we leave such analysis for future work.

Similarly, one could also combine the medium resolution eBOSS data with higher resolution \PoneD{} data from X-SHOOTER, HIRES or UVES, such as in \cite{Karacayli+21}. However, our post-processing and emulation technique has not yet been fully validated at $k > 0.05\,\mathrm{s km^{-1}}$ (where the main information of those datasets beyond eBOSS and DESI lies). We therefore opted to not yet combine with those high-resolution products.


\subsubsection{Cosmological prior}\label{ssec:planck_prior}
Since $\Omega_{\rm m} h^2$ cannot be measured in the data at high significance, and $H_0$ is unconstrained, we also check the impact of imposing a prior on $\{\Omega_{\rm m}, h\}$ from Planck 2018 data. This is shown in \cref{fig:data_fit_FP}, right panel. This prior has a strong impact mostly on $n_\lya$ due to the correlation between $n_\lya$ and $\Omega_{\rm m} h^2$ present in our constraints. Encouragingly, along the correlation between $\Omega_{\rm m}$ and $n_\lya$ the cosmological prior forces the constraints into a regime consistent with Planck data. The $A_\lya$ constraint is otherwise unchanged, whether we also include a prior on the mean transmitted flux or not. As such, we conclude that an informative prior on $\{\Omega_{\rm m}, h\}$ does not change any of our conclusions, and only slightly increases the constraining power in $n_\lya$.

When imposing both the mean transmission prior from \cite{Becker+13} as well as the prior on $\Omega_{\rm m} h^2$ from Planck, we find that we recover both an amplitude $A_\lya$ and a slope $n_\lya$ largely consistent with Planck data.

\section{Conclusions}\label{sec:conclusion}

\enlargethispage*{2\baselineskip}
In this work, we have introduced the \gridname{} suite of high resolution simulations performed with the \texttt{Nyx} hydrodynamical code. Each of these simulations followed $4096^3$ dark matter particles and baryon mesh cells in a 120Mpc comoving box ($\sim$ 81Mpc/$h$). These simulations are used together with a Gaussian process emulation technique, and the resulting emulator of the 1D flux power spectrum is subjected to a variety of validation tests. We perform leave-one-out tests, where a simulation is left out from the training set and the emulator is tasked to reproduce its flux power spectrum using only the information from the other simulations. The emulation accuracy is around 0.5-2\% in this test, depending on the redshift and wavenumber under consideration. We then check if fits using the full sampling pipeline consistently recovers a simulation that is not part of this training set (the ``fiducial'' simulation). The emulator nicely recovers the true cosmology without issues and this conclusion does not change when using an additional cosmological prior. We also fit the simulations described in \cite{Borde+13} and show that despite large differences in the codes and the simulation setup, the new emulator can still consistently recover the $A_\lya$ and $n_\lya$ parameters (with no sensitivity to $\Omega_{\rm m} h^2$) with only a mild bias on $A_\lya$ at the level of $-1.4\sigma$ without a cosmological prior and $-0.9\sigma$ with a cosmological prior.

\begin{table}[]
    \centering
    \bgroup
    \def\arraystretch{1.5}
    \setlength{\tabcolsep}{4pt}
    \begin{tabular}{c|c c | c c}
        Experiment & $A_\lya$ & $n_\lya$ & $\Delta_*$ & $n_*$ \\ \hline  \hline
        Planck (for reference) &  $8.79 \pm 0.17$ & $-2.3626 \pm 0.0031$ & $0.3476 \pm 0.0068$ & $-2.2899 \pm 0.0032$ \\ \hline
        \gridname{} only & $< 7.6$ & ${-2.3693}^{+0.0082}_{-0.0074}$ & $0.260 \pm 0.024$ & ${-2.2995}^{+0.0068}_{-0.0064}$\\ \hline
        \multirow{2}{*}{\begin{tabular}{c}\gridname{} + $\tau_\mathrm{eff}(z)$ \\[-5pt]+ $\{\Omega_m , H_0\}$ priors\end{tabular}} & \multirow{2}{*}{$9.8 \pm 1.1$} & \multirow{2}{*}{$-2.3705 \pm 0.0068$} & \multirow{2}{*}{$0.388 \pm 0.045$} & \multirow{2}{*}{$-2.2978 \pm 0.0067$}  \\
        & & & & 
    \end{tabular}
    \egroup
    \caption{Main results of this work: Mean and 68\% credible intervals (or 95\% upper limit) for the parameters representing the amplitude and slope of the linear power spectrum at the scales probed by the Lyman-$\alpha$ forest. The constraints are illustrated in \cref{fig:moneyplot}.\label{tab:results}}
\end{table}
\begin{figure}
    \centering
    \includegraphics[width=0.66\linewidth]{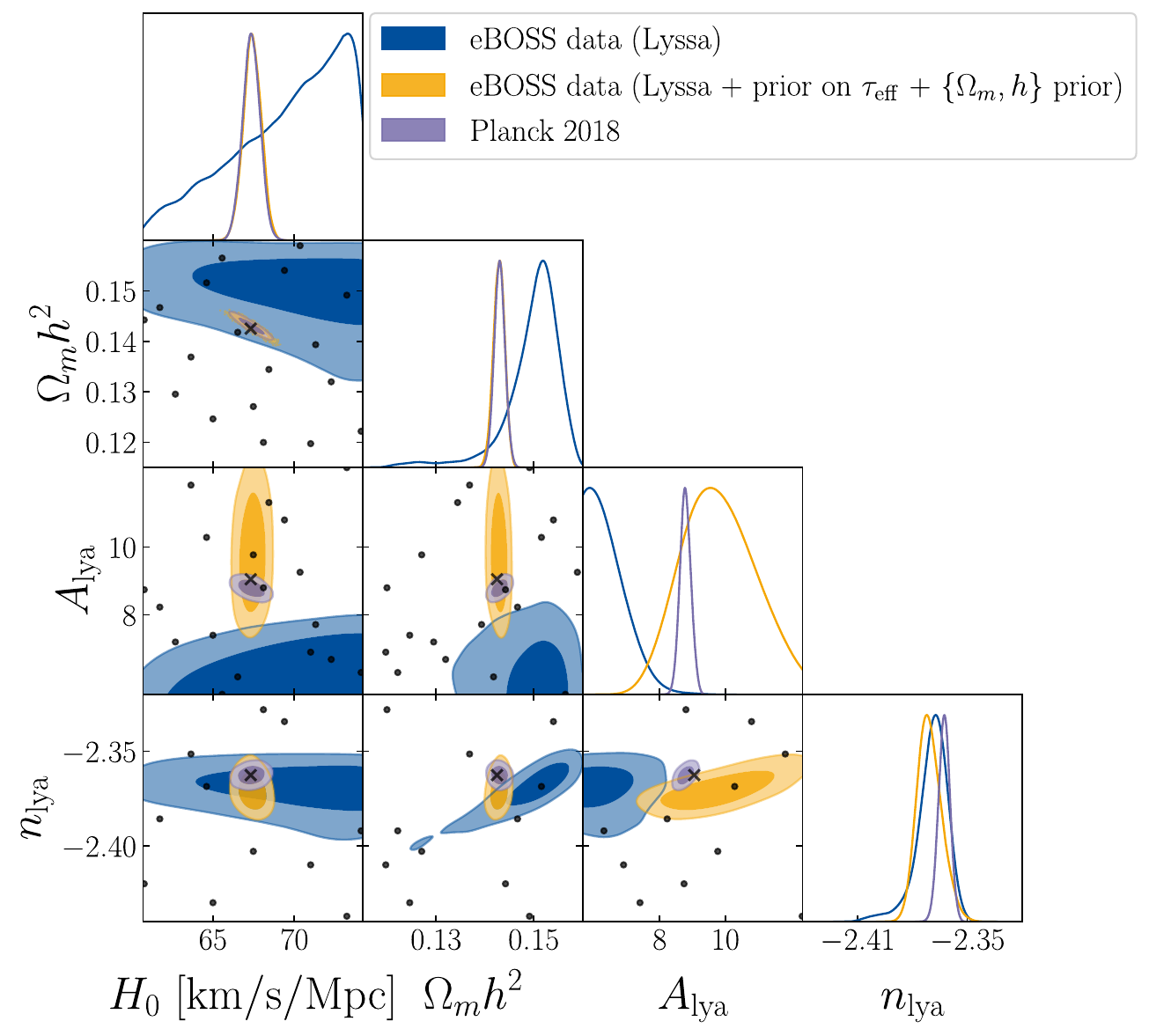}
    \includegraphics[width=0.33\linewidth]{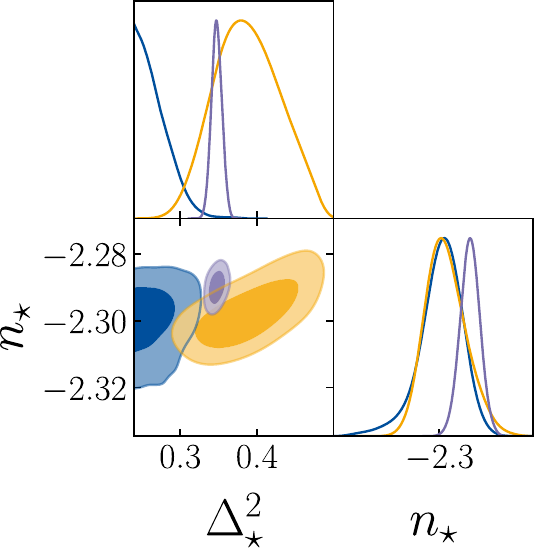}
    \caption{Same as the right panel of \cref{fig:data_fit_FP}, but showing only the case without any additional informative priors and the case with a prior on $\{\Omega_m,h\}$ and $\tau_\mathrm{eff}(z)$\,. Left: Cosmological part of the parameter basis. Right: Translation into the $\{\Delta_*, n_*\}$ parameter space (using an emulator technique). Notice the similarity to the $\{A_\lya, n_\lya\}$ parameter space. The numerical values are summarized in \cref{tab:results}.}
    \label{fig:moneyplot}
\end{figure}
We further apply the emulator and fitting pipeline to the 1D Lyman-$\alpha$ data from eBOSS DR14 derived in \cite{Chabanier+19}. We find a tension in the amplitude $A_\lya$ compared to post-processed constraints from Planck 2018 (from \cite{Planck:2018vyg}). This tension is reduced slightly when only data at $z\geq 2.6$ are considered (as advocated in \cite{Fernandez+23}). More significantly, we find that such a tension is largely driven by a degeneracy between the amplitude $A_\lya$ and the mean transmission of the Lyman-$\alpha$ forest at $z=3$ ($A_\tau$), whose best-fit value lies towards the edge of the prior in the baseline analysis.\enlargethispage{-2\baselineskip}

On this basis, we argue that the inclusion of a mean transmission prior from \cite{Becker+13} is crucial to keep the \gridname{} emulator in a well-behaved regime. With this well-motivated prior we find that our inferred value for $A_\lya$\, becomes completely consistent with the results from Planck 2018, and our inferred parameters do not show any large changes when fitting only $z\geq 2.6$. We thus note that further investigations into systematic uncertainties based on the modeling of the mean flux transmission as well as into the differences between the emulation techniques will be necessary to improve the robustness of cosmological constraints from the 1D Lyman-$\alpha$ forest. We summarize our results in \cref{fig:moneyplot,tab:results}.

Further comparisons with other simulation suites, such as those of \cite{Bird+23} will establish the consistency of the conclusions. The simulations presented here may also be used in future \PoneD{} analyses in the context of DESI and for the analysis of the three-dimensional flux transmission power spectrum, though we expect that the employed emulation techniques will have to be adapted again for each case.

\begin{acknowledgments}

The authors would like to thank Andreu Font-Ribera, Jonas Chaves-Montero, and Laura Cabayol-Garcia from the IFAE group for several useful discussions, as well as the DESI Lya WG for their continued support and interesting ideas. We would like to thank Fajr Abdulrahman for his help with the conversion between $\{A_\lya\,, n_\lya\}$ and $\{\Delta^2_*\,, n_*\}$. The authors would also like to thank Felix Kahlh{\"o}fer for useful discussions. We finally thank Simeon Bird for comments on an earlier version of this paper.

We acknowledge PRACE for awarding us access to Joliot-Curie at TGCC, France via proposal 2019204900. We acknowledge the French national access to high-performance computing GENCI (Grand Equipement National de Calcul Intensif), for additional access to the HPC resources of TGCC under the allocations A0050410586, A0070410586 and AD01413571. 

We also acknowledge the Oak Ridge Leadership Computing Facility, which is supported by the Office of Science of the U.S.~Department of Energy under Contract No.~DE-AC05-00OR22725. This work used resources of National Energy Research Scientific Computing Center (NERSC), a U.S.~Department of Energy Office of Science User Facility located at Lawrence Berkeley National Laboratory, operated under Contract No.~DE-AC02-05CH11231. MW acknowledges support by the project
AIM@LMU funded by the German Federal Ministry of Education
and Research (BMBF) under the grant number 16DHBKI013. NS acknowledges the support of the following Maria de Maetzu fellowship grant: Esta publicaci\'on es parte de la ayuda CEX2019-000918-M, financiada por MCIN/AEI/10.13039/501100011033. NS acknowledges support by MICINN grant number PID 2022-141125NB-I00. NS also was supported by a Fraunhofer-Schwarzschild Fellowship at Universitäts Sternwarte München. JL and MM acknowledge support from the DFG grant LE 3742/8-1. EA, CY and NPD acknowledge support from the ANR grant ANR-22-CE92-0037.\enlargethispage{4\baselineskip}

This work was partially supported by the DOE’s Office of Advanced Scientific Computing Research and Office of High Energy Physics through the Scientific Discovery through Advanced Computing (SciDAC) program. The development of \texttt{Nyx} as an AMReX application was supported by the U.S. Department of Energy, Office of Science, Office of Advanced Scientific Computing Research, Applied Mathematics program under contract number DE-AC02005CH11231, and by the Exascale Computing Project (17-SC-20-SC), a collaborative effort of the U.S. Department of Energy Office of Science and the National Nuclear Security Administration.

\end{acknowledgments}

\bibliographystyle{JHEP}
\bibliography{biblio,software}

\appendix
\clearpage

\begin{table}[h]
    \centering
    \resizebox{\textwidth}{!}{%
    \begin{tabular}{r|c c | l}
        Parameter &  Minimum & Maximum & Description\\ \hline
        $T_0(z=3)$ & 0 & 25000  & Temperature at mean density at $z=3$ in K [see \cref{eq:powerlaws}]\\
        $\eta_{T}$ & -5 & 10 &Logarithmic slope of $T_0$ at $z<3$ [see \cref{eq:powerlaws}]\\
        $\Delta \eta_{T} $ & -15 & 12  & Difference in logarithmic slope between $z<3$ and $z>3$ [see \cref{eq:powerlaws}]\\
        $\gamma(z=3)$ & 0.3 & 2.0  & Logarithmic slope of temperature-density relation at $z=3$\\
        $\eta_{\gamma}$ & -5 & 2  & Logarithmic slope of $\gamma$ at $z<3$\\
        $A_\tau$ & 0 & 1.5  & Effective optical depth at $z=3$ [see \cref{eq:powerlaws}]\\
        $\eta_\tau$ & 0 & 7.0  & Logarithmic slope of effective optical depth $\tau$ [see \cref{eq:powerlaws}]\\
        $\mathbf{A_{AGN}}$ & 0 & 3& AGN correction amplitude of \cref{eq:agn}\\
        $\mathbf{A_{SN}}$ & 0 & 3  & SN correction amplitude of \cref{eq:sn}\\ 
        $\mathbf{A_{n,i}}$ & 0 & 3  & Amplitude of noise power correction of bin $i$ \\ 
        $f_\mathrm{SiIII}$ & - 0.2 & 0.2  & Amplitude $a$ of \cref{eq:metal} for SiIII scaled inversely with mean flux \\
        $f_\mathrm{SiII}$& - 2.0 & 2.0  & Amplitude $a$ of \cref{eq:metal} for SiII scaled inversely with mean flux\\
        \hline
        $\lambda_P$ & 60 & 100 & Pressure smoothing scale at $z=3$\\
        $\eta_{\lambda_P}$ & -50 & 50 & Logarithmic slope of $\lambda_P$ at $z<3$\\
        $\Delta \eta_{\lambda_P}$ & -50 & 50 & Difference in logarithmic slope of $\lambda_P$ between $z<3$ and $z>3$\\ \hline
        $\mathbf{A_{UVfluct}}$ & 0 & 3 & UV fluctuation amplitude\\
    \end{tabular}}
    \caption{Maximum allowed prior ranges on astrophysical/nuisance parameters. These are similar to those of \cite{Palanque+19}, except for the changes explicitly mentioned in the text. There are 13 noise parameters $A_{n,i}$ with $i \in [1..13]$, one for each redshift bin. Additional Gaussian priors are described for the bold parameters in \cref{ssec:nuisances}. Note that there are additional effective priors from the finite extent of the \gridname{}-based emulator, which are shown in \cref{fig:app:implicit}.}
    \label{tab:priors}
\end{table}

\begin{figure}
    \centering
    
    \hspace{0.2\textwidth}\includegraphics[width=0.247\textwidth]{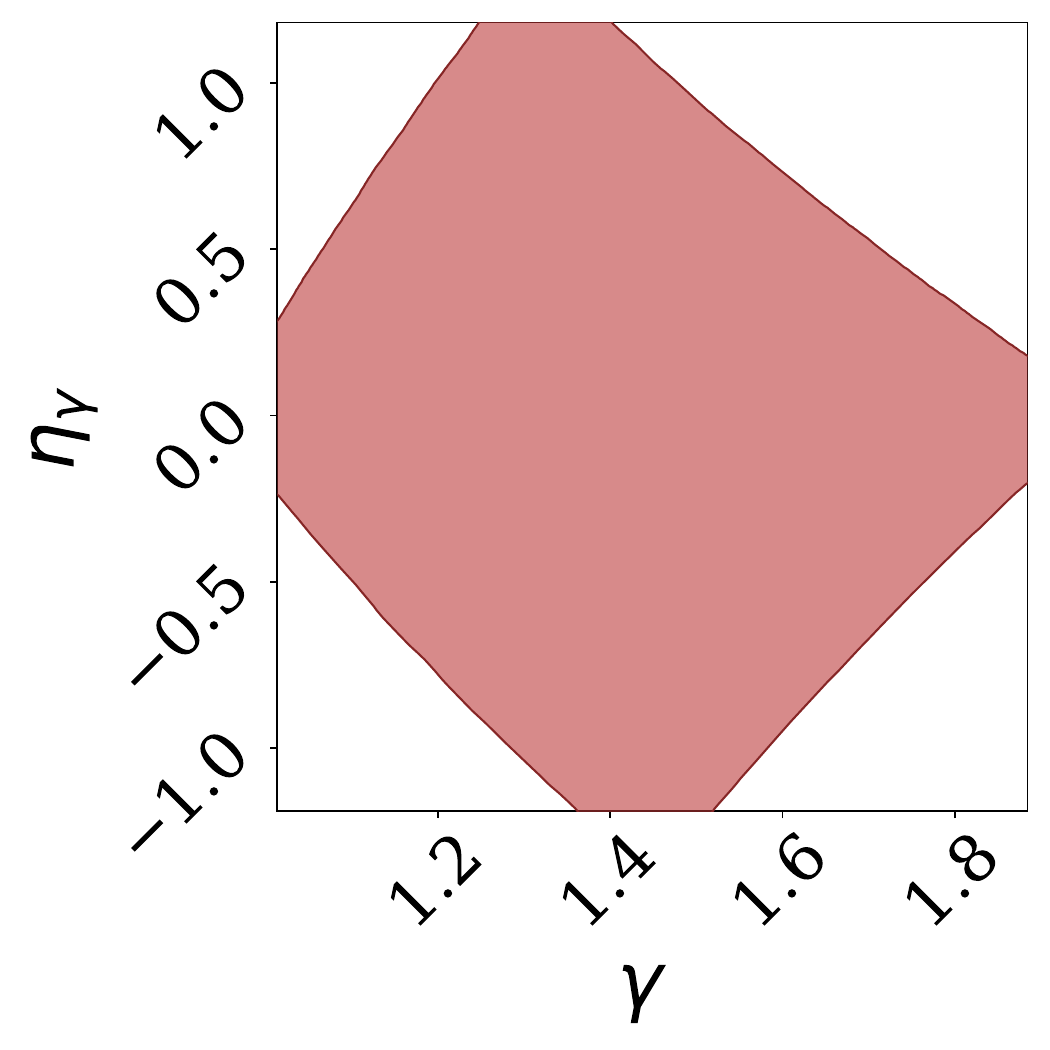}\hspace{0.25\textwidth}
    \includegraphics[width=0.247\textwidth]{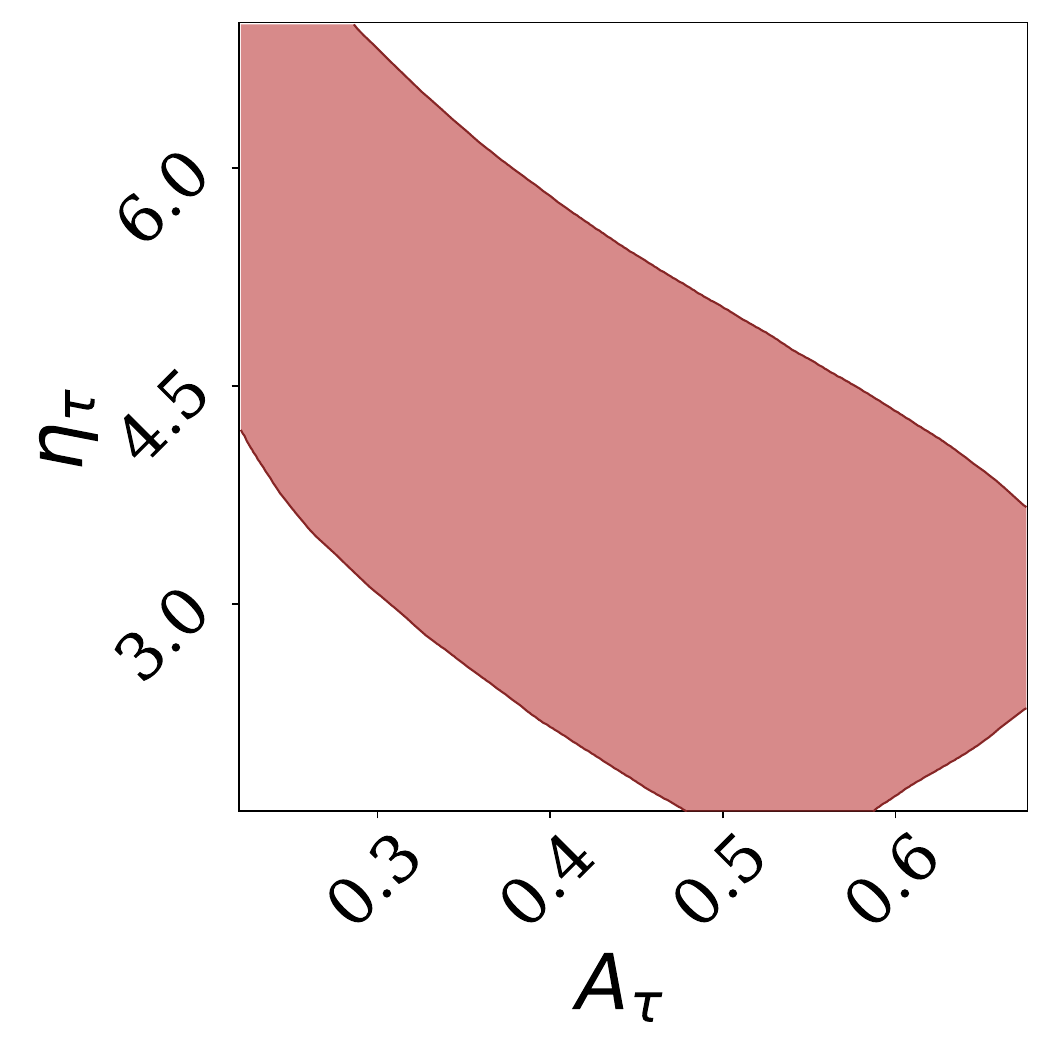}\\
    \includegraphics[width=0.45\textwidth]{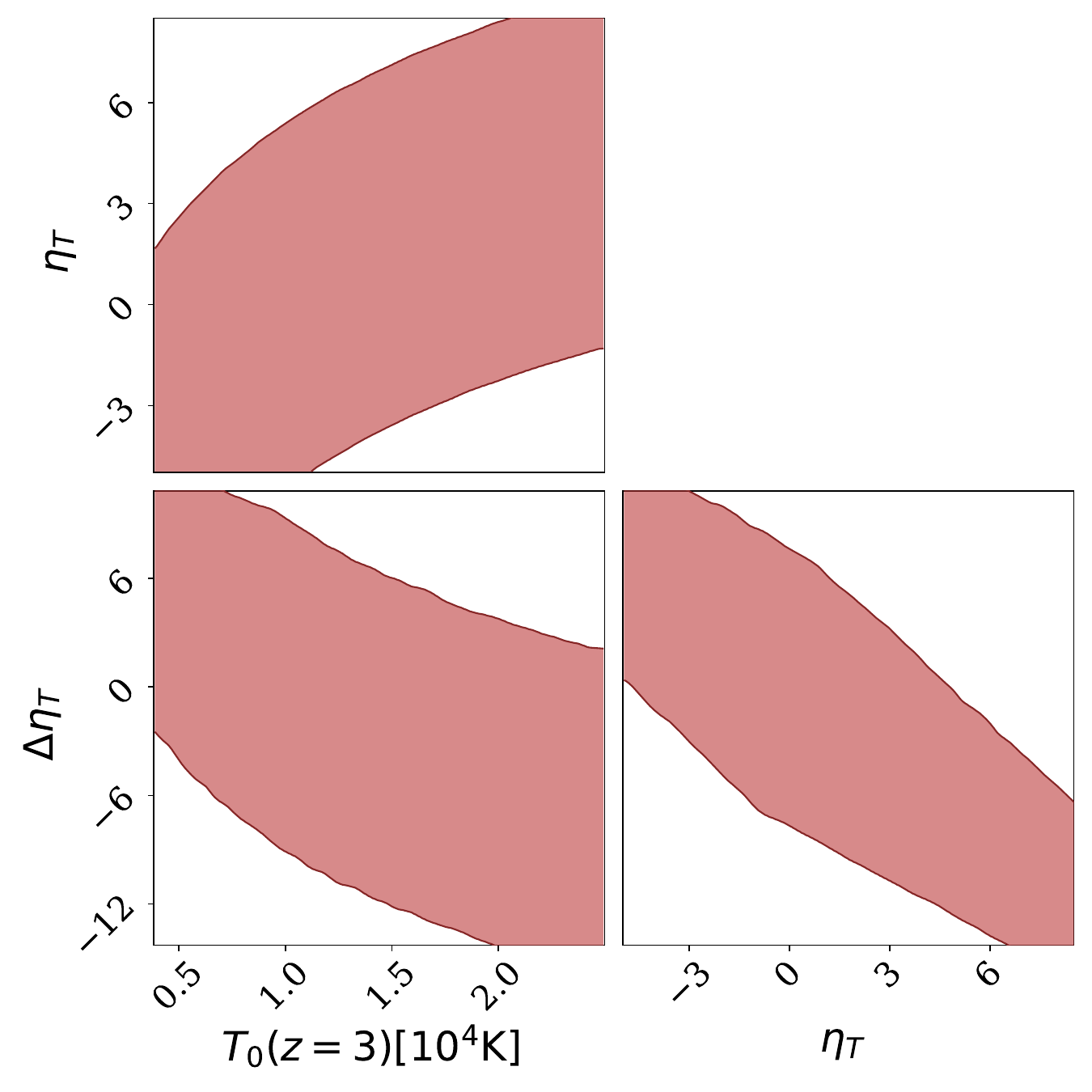}\hspace{0.05\textwidth}
    \includegraphics[width=0.45\textwidth]{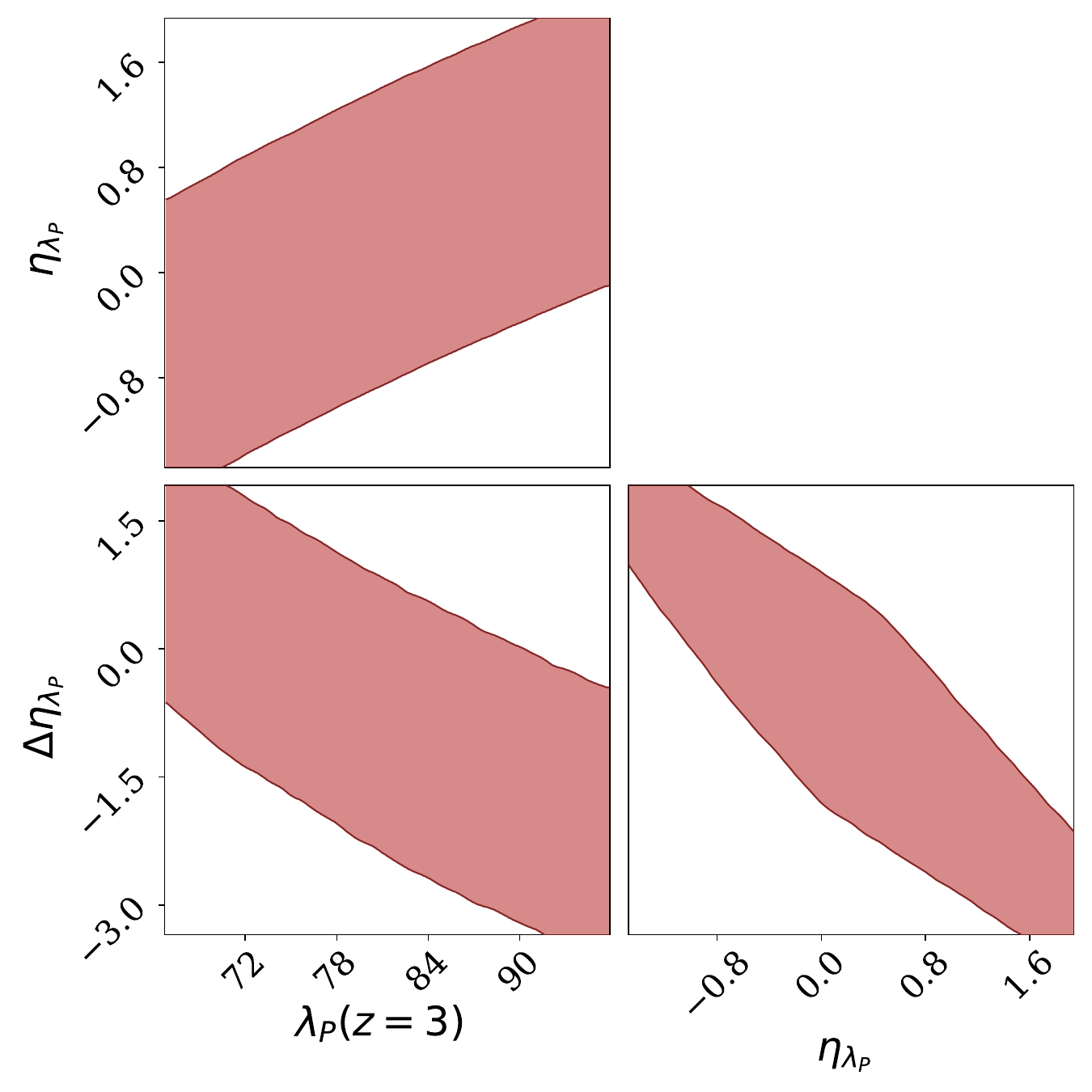}

    \caption{\label{fig:app:implicit}Implicit priors set on thermal parameters for all constraints involving the \gridname{} simulations with redshift evolution of parameters being described by (broken) power laws. Contours show the ranges reached when sampling only the priors (explicit from \cref{tab:priors} and implicit from the design of simulations) for all relevant 2d-subspaces. Priors are uniform inside the red areas for the parameters shown, but zero outside. Note that for fits with free $\bar{F}$ (free $\lambda_P$) evolution, the implicit priors for $A_\tau,\eta_\tau$ ($\lambda_P$, $\eta_{\lambda_P}$, $\Delta \eta_{\lambda_P}$) do not apply.
    }
\end{figure}
\section{Description of nuisance parameters}\label{ssec:app:nuisance}
In \cref{tab:priors} we show the priors on the parameters that we used during the analysis. Note that these priors were only used when the parameters were not fixed (as they were for the validation in \cref{ssec:l1o,ssec:fiducial,ssec:previous}). Also that most of these parameters are equivalent to those described in \cite{Palanque+19}, apart from a few exceptions. For example, instead of modeling the temperature slopes at $z<3$ and $z>3$ with different parameters, we instead put a linear prior on the difference $\Delta \eta^{T_0} = \eta^{T_0}(z>3) - \eta^{T_0}(z<3)$ between the slopes. Additionally, we use a power law parameterization for the pressure smoothing scale $\lambda_P$ instead of a simple overall rescaling of the heating rate. We put additional Gaussian priors on the noise level ($0 \pm 0.02$ for each redshift bin), the SN correction amplitude ($1 \pm 0.3$), the AGN correction amplitude ($1 \pm 0.3$), and for the Taylor emulator also on the UV fluctuation amplitude ($0 \pm 0.3$) and the splicing parameters as in \cite{Palanque+19} (offset $0.01 \pm 0.05$ and slope $0.9\pm 5.0$).

Note that due to the finite extent of the \gridname{} simulation thermal reprocessings, there are effective thermal priors on the powerlaw parameter basis, which are shown in \cref{fig:app:implicit}.

\section{Variations of the baseline analysis\label{app:ssec:free_thermal}}

\begin{figure}
    \centering
    \includegraphics[width=0.49\textwidth]{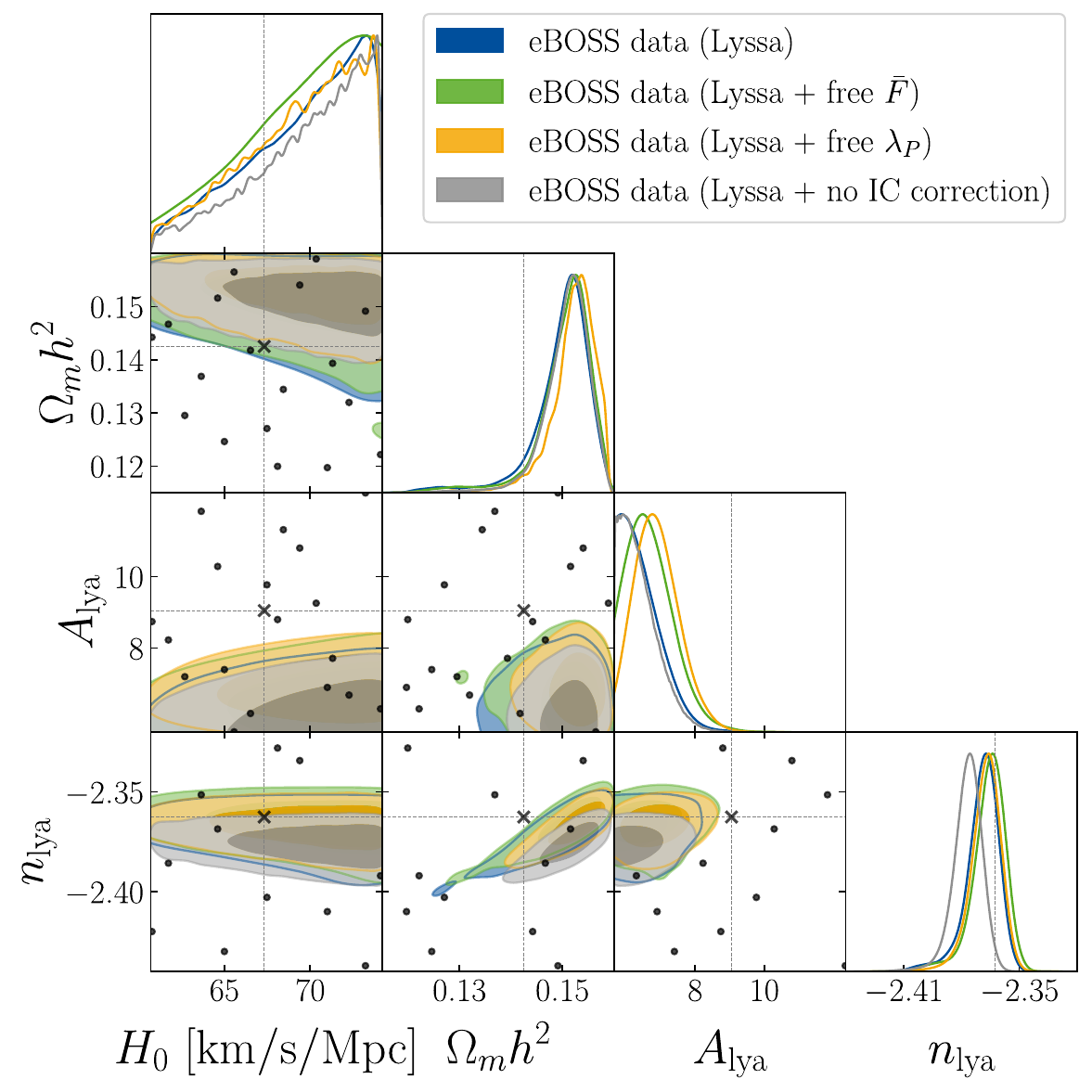}
    \caption{Results equivalent to \cref{fig:data_fit} obtained when leaving the mean transmission ($\bar{F}$) redshift dependence free, when leaving the pressure smoothing scale ($\lambda_P$) redshift dependence free, or when turning off the IC correction.}
    \label{fig:variations}
\end{figure}
In this section we present variations of the baseline analysis (\cref{fig:data_fit}) in which we allow additional freedom to the fit (without imposing additional informative priors). See \cref{fig:variations}. The \enquote{free $\bar{F}$} case allows a free mean transmission redshift dependence (with a nuisance parameter for each redshift bin and an implicit prior set by the range of our simulation grid at that redshift, see \cref{ssec:app:nuisance}), whereas the \enquote{free $\lambda_P$} allows a free $\lambda_P$ redshift dependence (again limited by the extent of our model grid). Finally, the \enquote{no IC} case tests what happens if the IC correction of \cref{ssec:ic} is disabled. In none of these cases we find a significant relaxation of the tension in $A_\mathrm{lya}$\,.

\section{Best-fitting power spectra}
\begin{figure}
    \centering
    \includegraphics[width=0.49\linewidth]{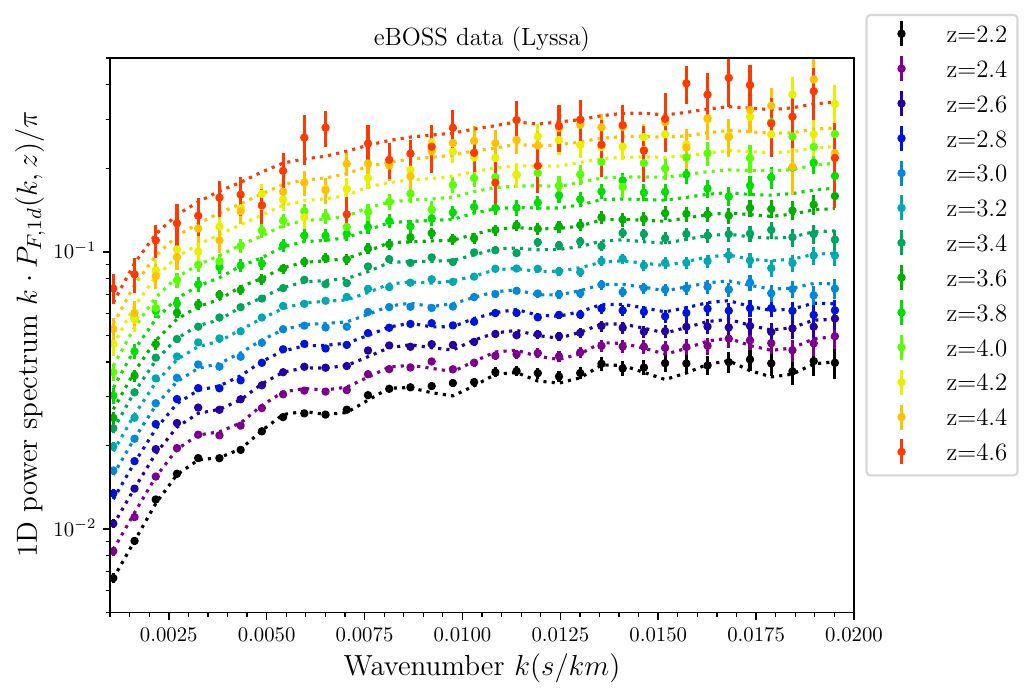}
    \includegraphics[width=0.49\linewidth]{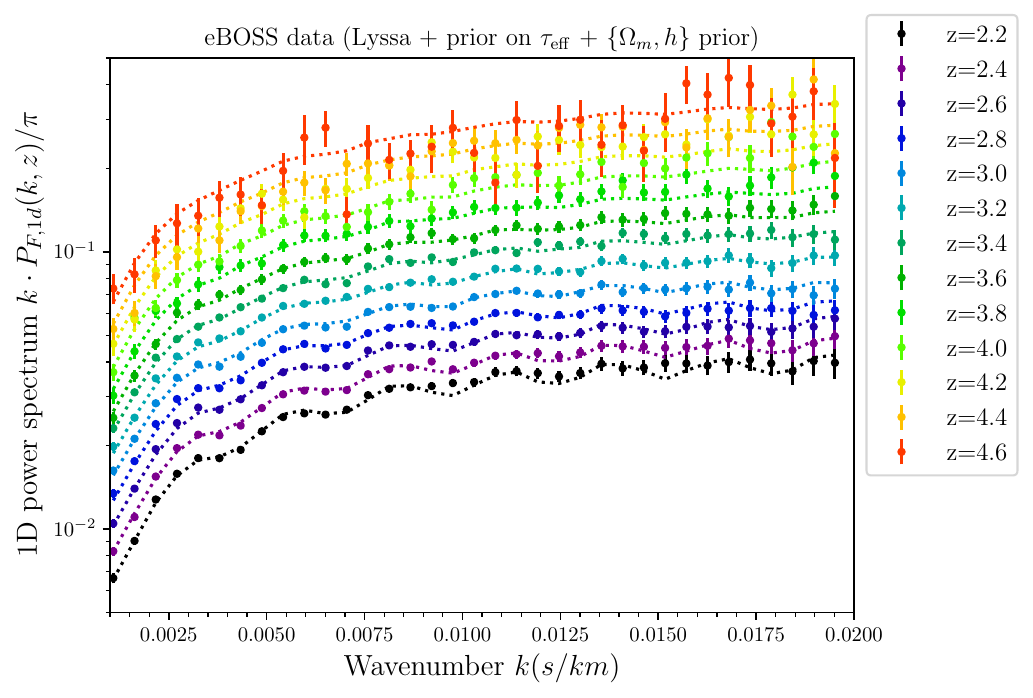}
    \caption{Best-fitting models compared to the data for the two cases shown in \cref{fig:moneyplot}. The data is the same as in \cite{Palanque+19}, and the displayed theoretical predictions include the nuisance parameter corrections described in \cref{ssec:nuisances}.}
    \label{fig:bestfit}
\end{figure}

We show the best-fitting power spectra for the main results of the eBOSS analysis from \cref{ssec:eBOSS_data} in \cref{fig:bestfit}. Both our baseline results and the results including a cosmological and effective mean transmission prior are presented. We note that the fit in both cases looks very good and are comparable to that of the bottom panel of Figure 1 of \cite{Palanque+19}.
\FloatBarrier
\section{Additional plots of constraints on nuisance parameters}

\begin{figure}
    \centering
    \includegraphics[width=0.84\textwidth]{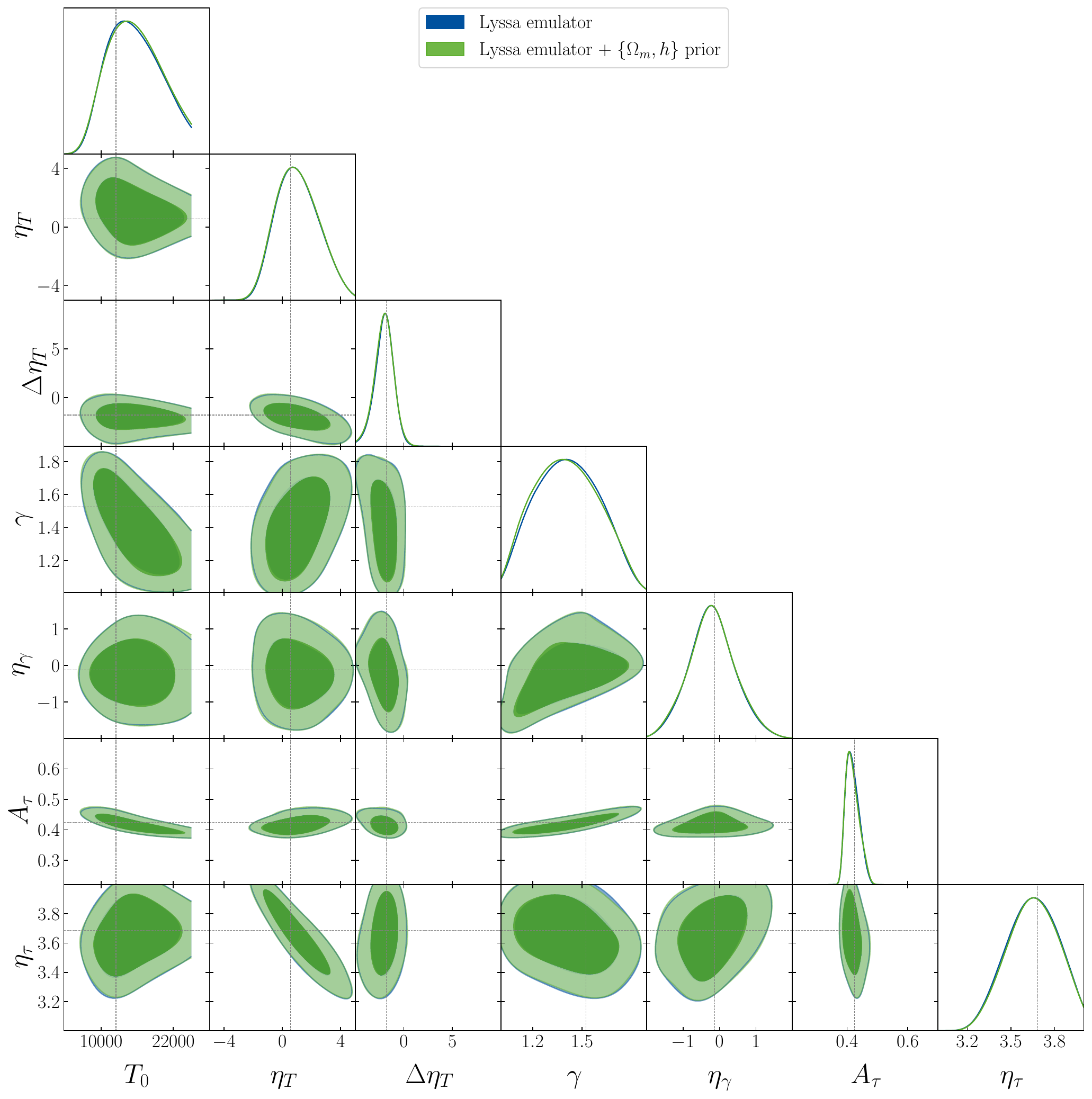}\\
    \includegraphics[width=0.84\textwidth]{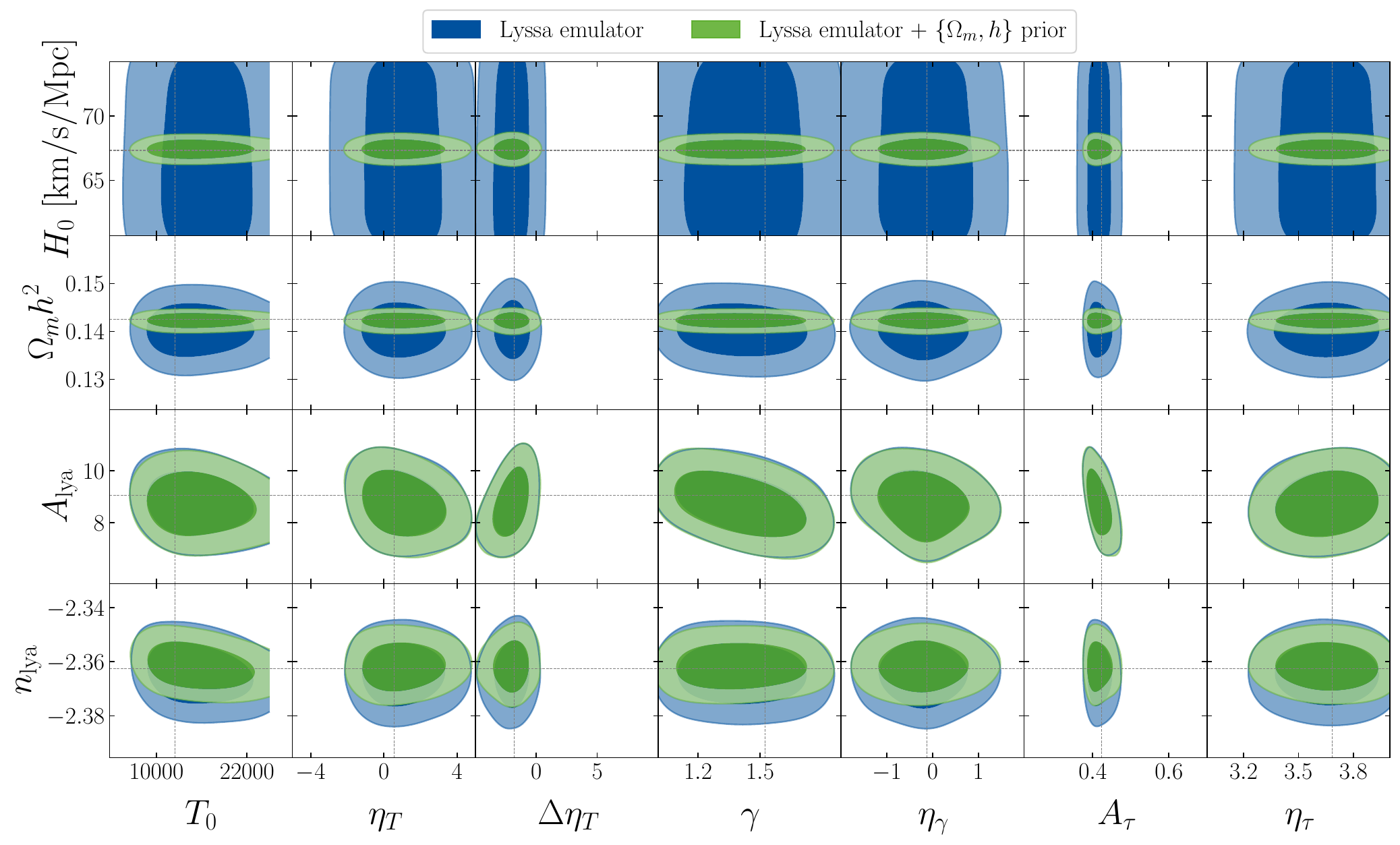}
    \caption{The 68\% and 95\% CL bounds for nuisance parameters when fitting the fiducial simulation with the \gridname{}-based emulator, corresponding to \cref{fig:fid_fit}.}
    \label{fig:fid_fit_nuisance}
\end{figure}

\begin{figure}
    \centering
    \includegraphics[width=0.84\textwidth]{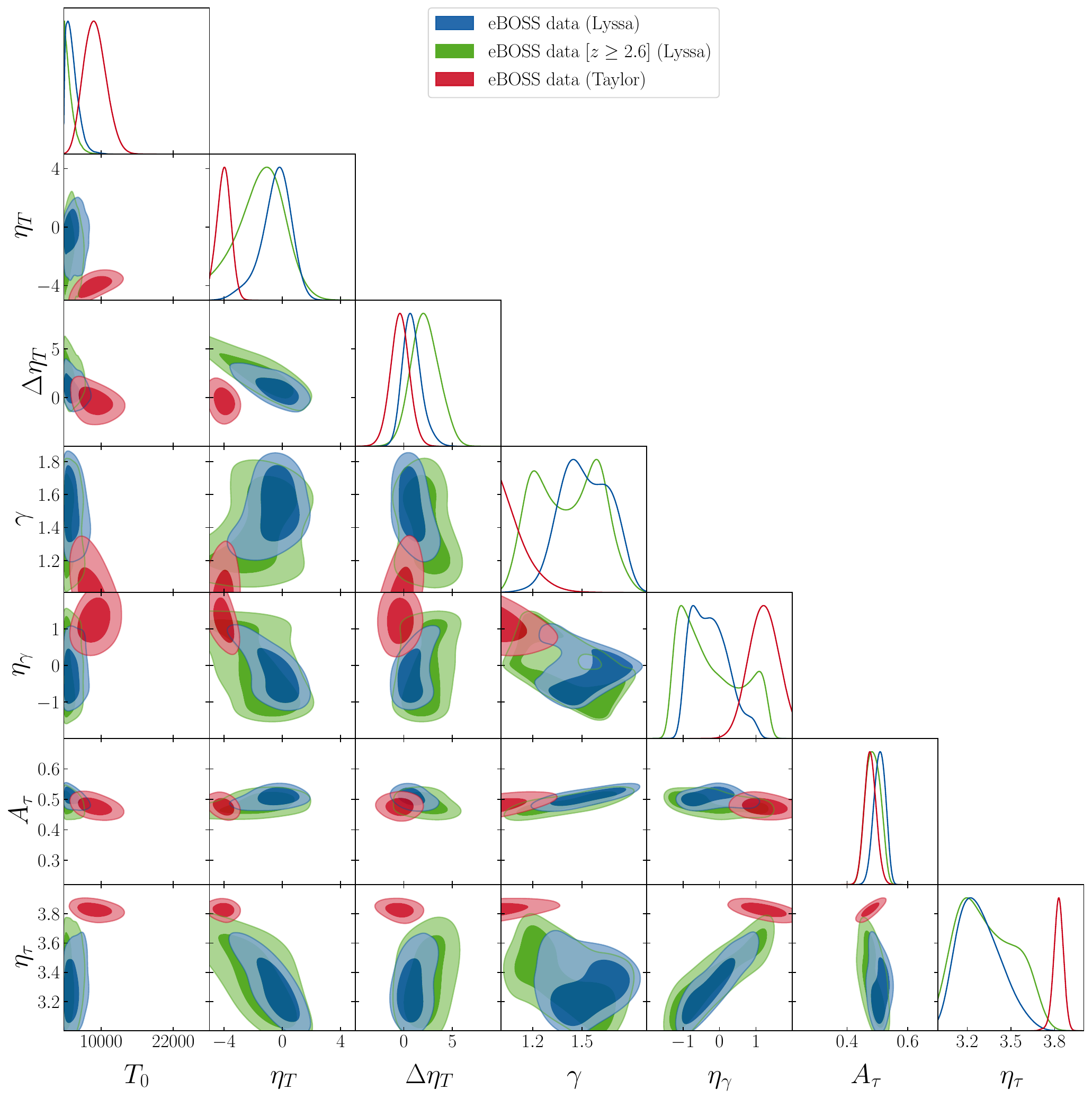}\\
    \includegraphics[width=0.84\textwidth]{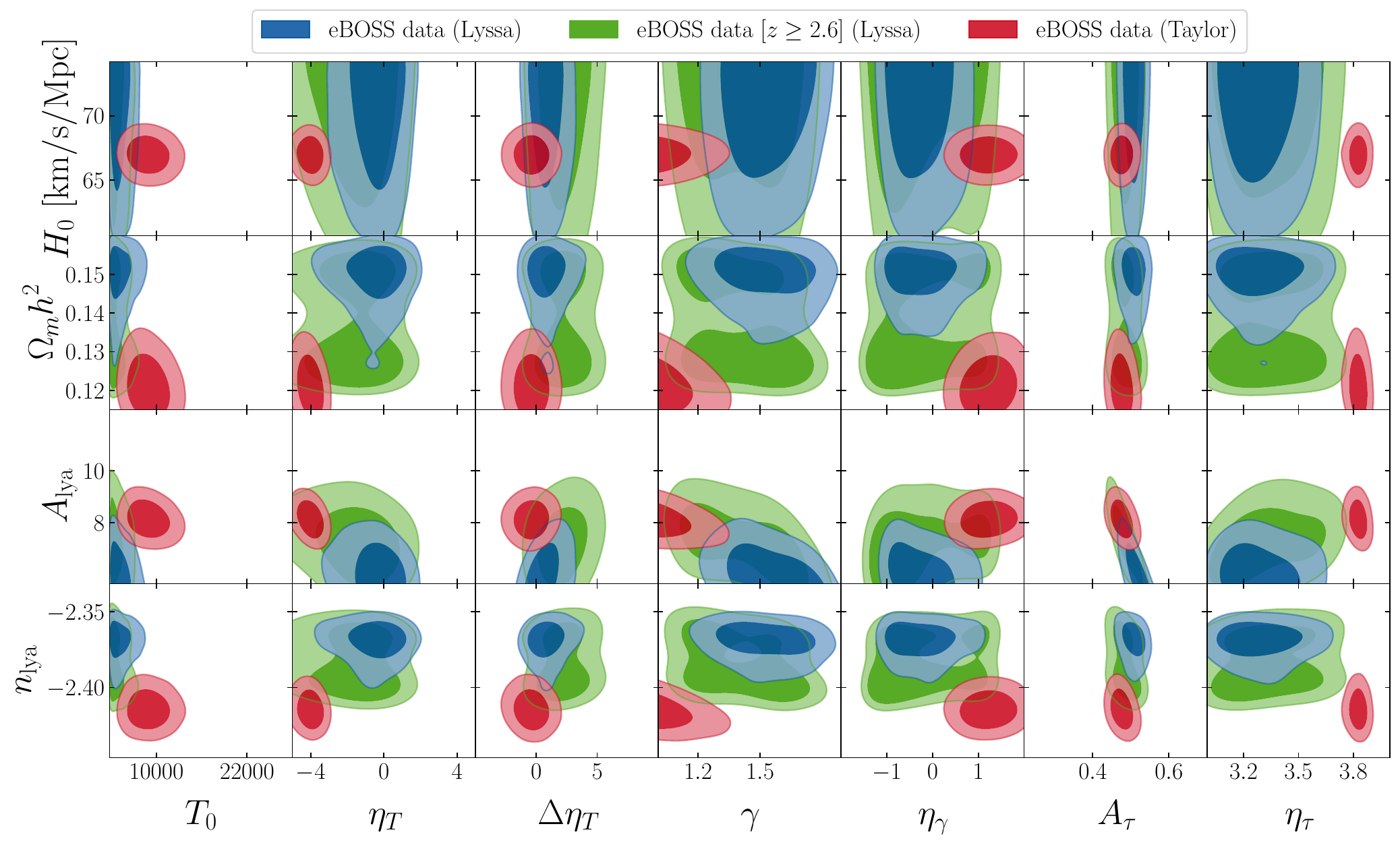}
    \caption{Same as \cref{fig:fid_fit_nuisance}, but instead corresponding to the run of \cref{fig:data_fit} (left panel).}
    \label{fig:data_fit_nuisance}
\end{figure}

In this section we present the correlations of the cosmological parameter constraints with those of the astrophysical and nuisance parameters. These are shown in \cref{fig:fid_fit_nuisance} for the fit to the fiducial simulation of \cref{ssec:fiducial} and in \cref{fig:data_fit_nuisance} for the baseline fit to the eBOSS data. If lines are shown in the corresponding plots for fit to the fiducial simulation, they show the thermal parameters that would correspond to the best-fitting (broken) power law models corresponding to the true thermal history of the fiducial simulation as well as the true cosmological parameters of that simulation. For example, we fit a broken powerlaw in to the actual $T_0(z)$ of the fiducial simulation and derive the corresponding $\{T_0 , \eta_T, \Delta \eta_T \}$ parameter set.

\begin{figure}
    \centering
    \includegraphics[width=0.84\textwidth]{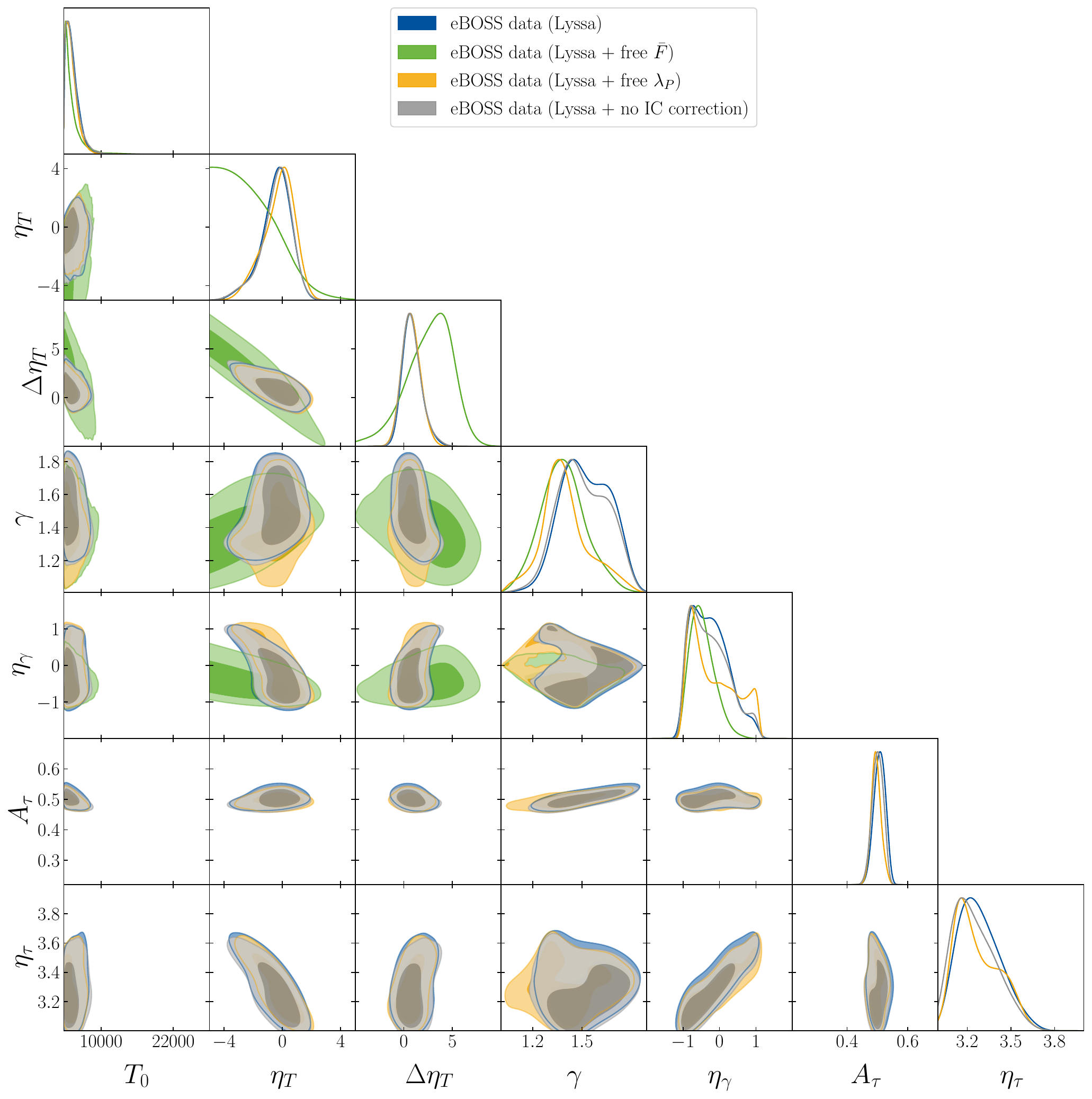}\\
    \includegraphics[width=0.84\textwidth]{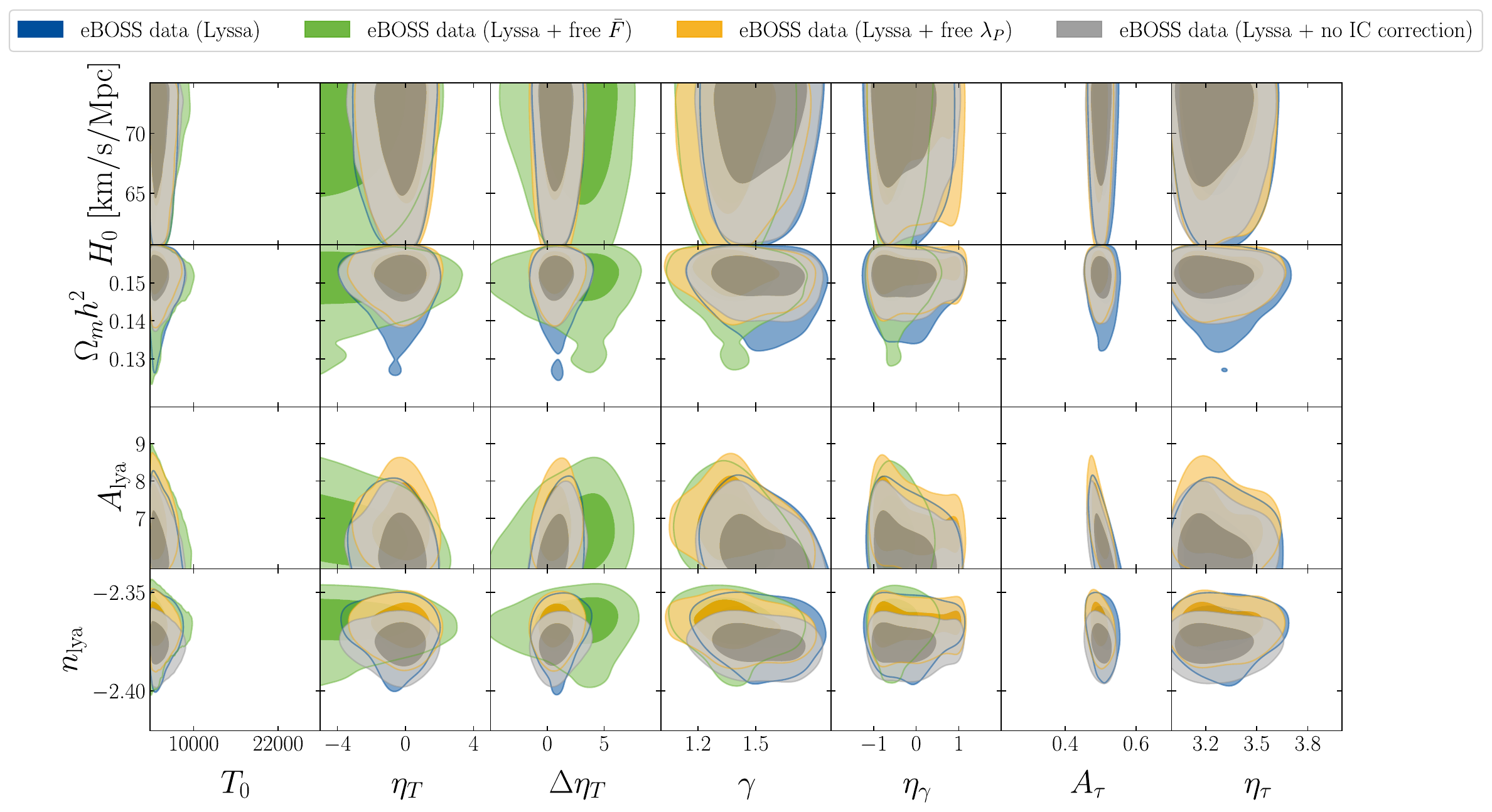}
    \caption{Same as \cref{fig:fid_fit_nuisance}, but instead corresponding to the run of \cref{fig:data_fit} (right panel).}
    \label{fig:data_fit_nuisance_2}
\end{figure}

\begin{figure}
    \centering
    \includegraphics[width=0.84\textwidth]{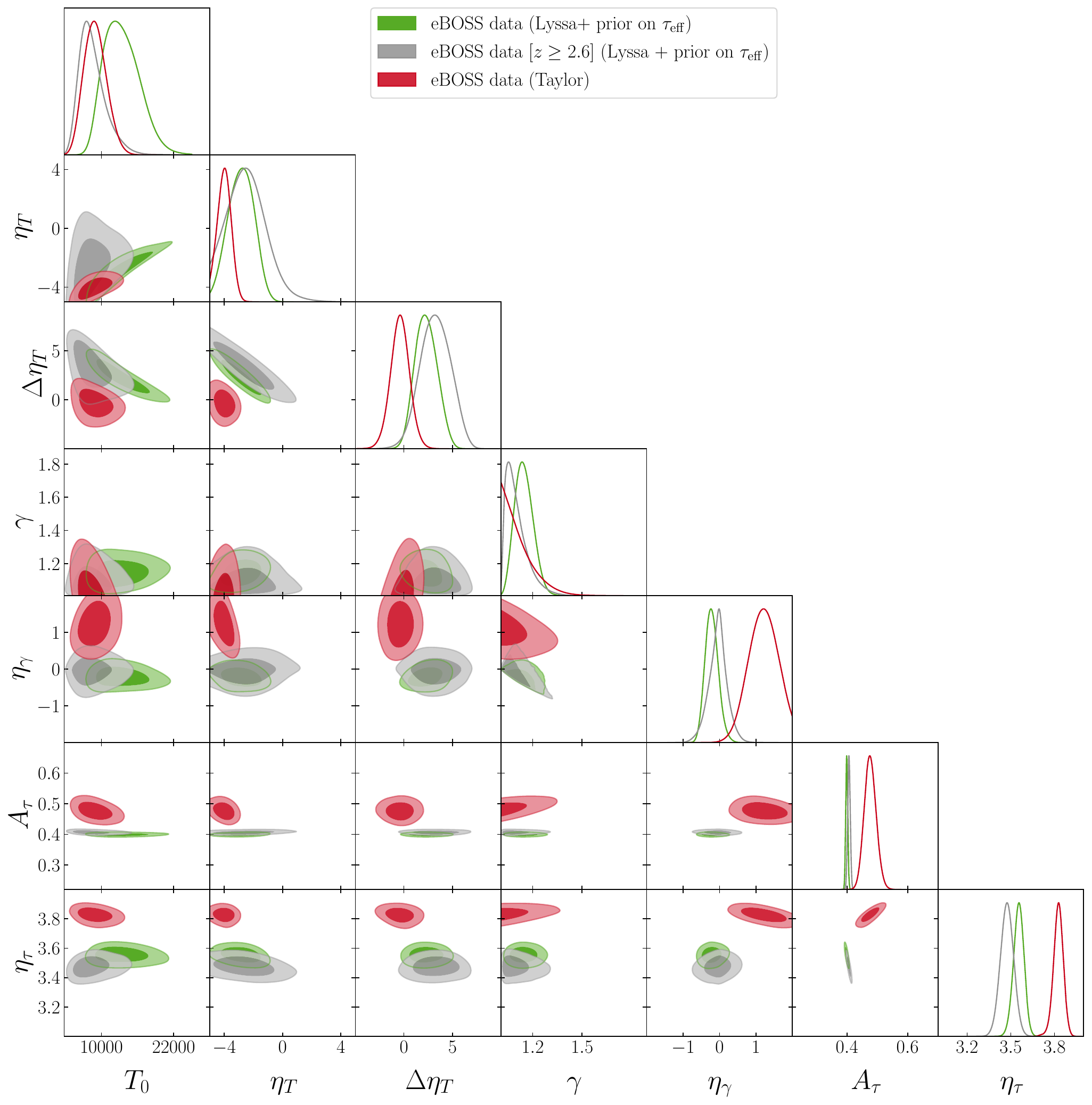}\\
    \includegraphics[width=0.84\textwidth]{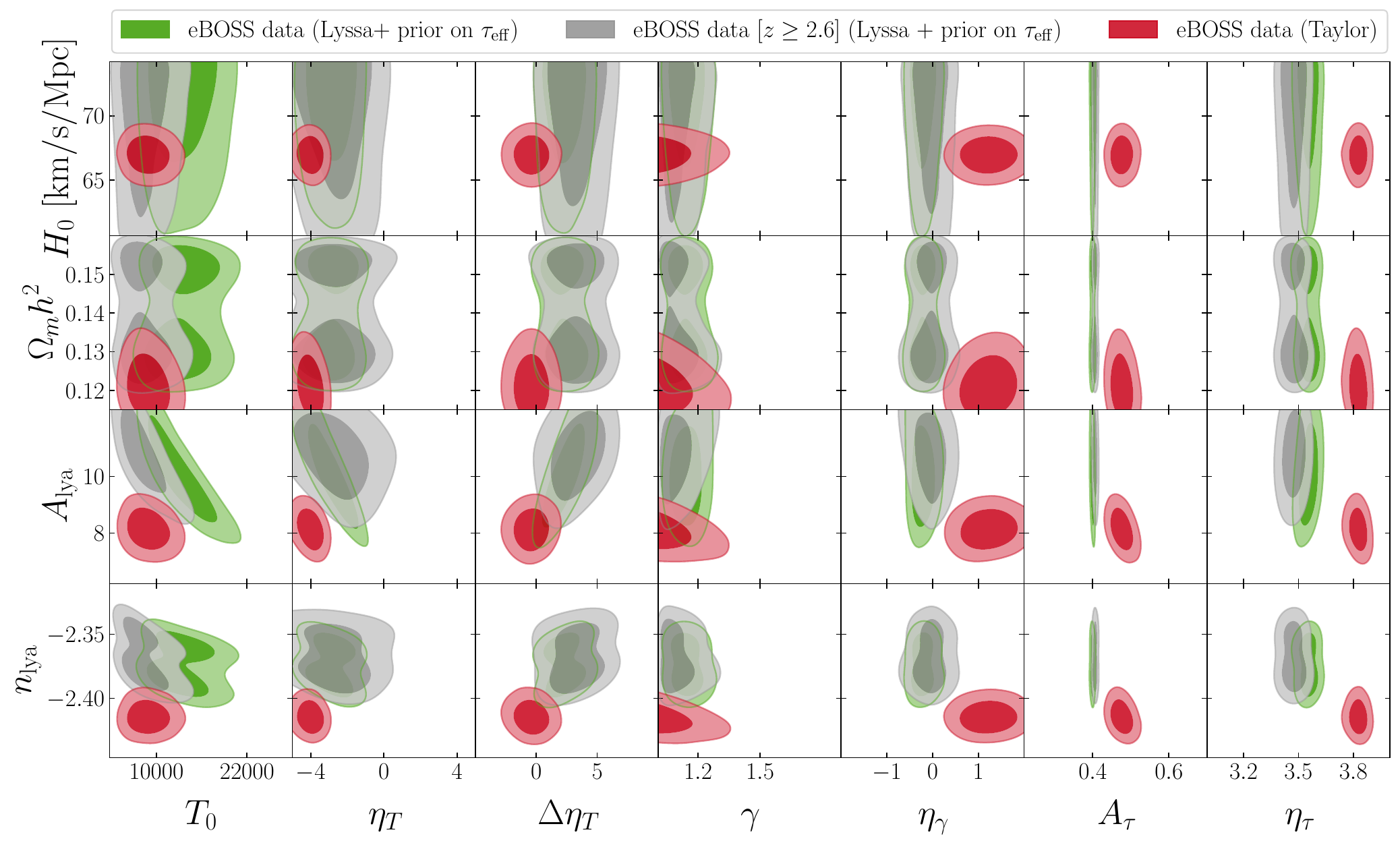}
    \caption{Same as \cref{fig:fid_fit_nuisance}, but instead corresponding to the run of \cref{fig:data_fit_FP} (left panel).}
    \label{fig:data_fit_FP_nuisance}
\end{figure}

\begin{figure}
    \centering
    \includegraphics[width=0.84\textwidth]{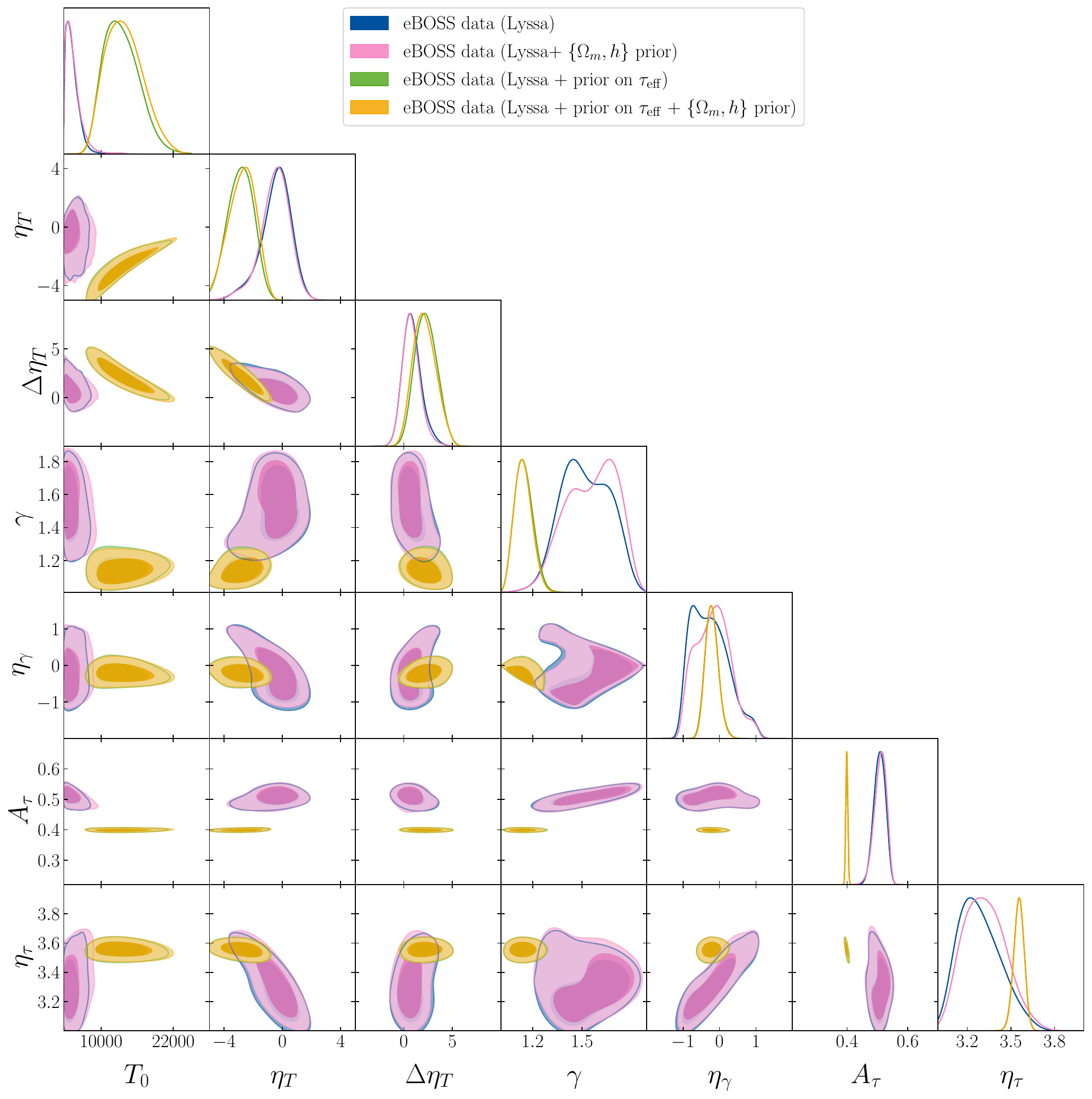}\\
    \includegraphics[width=0.84\textwidth]{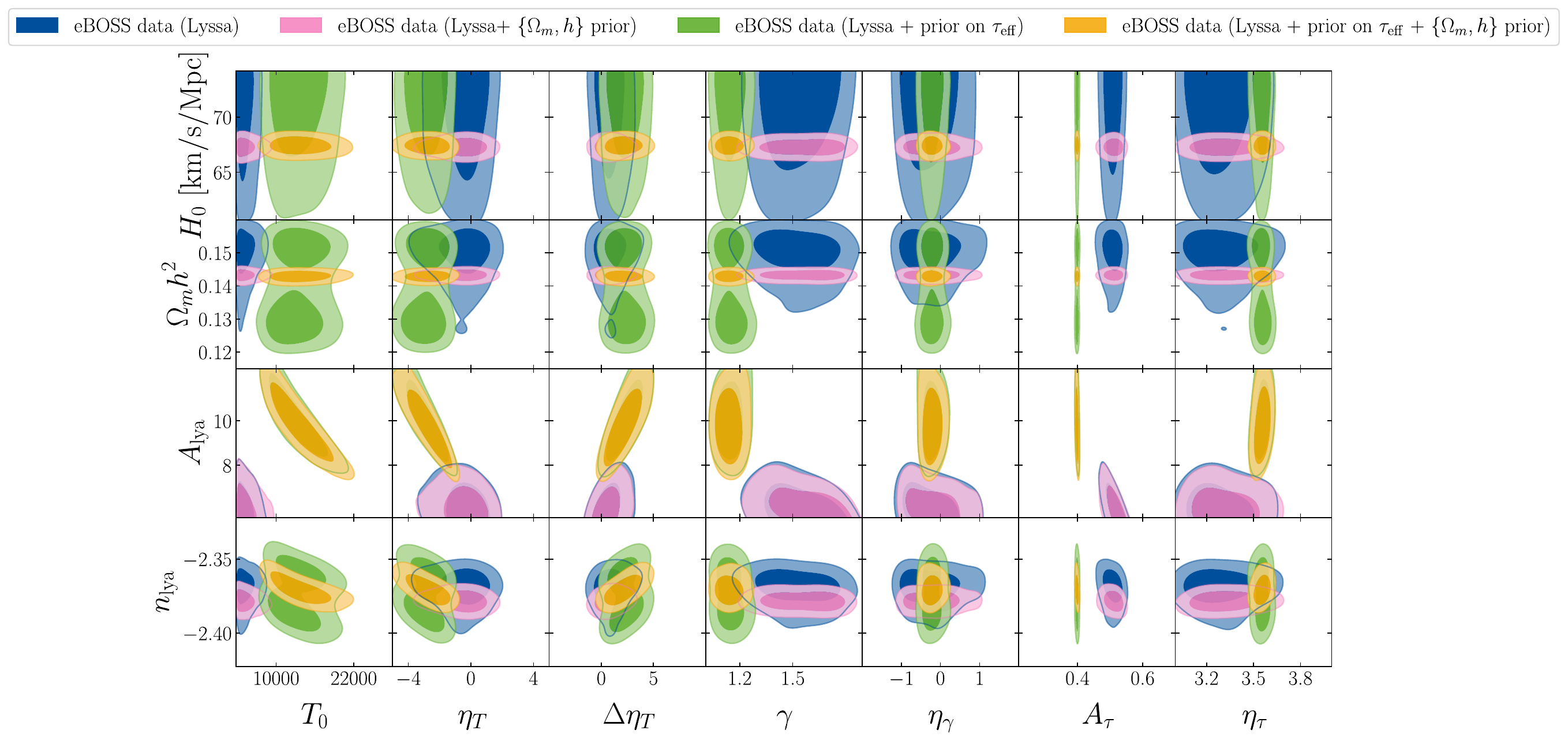}
    \caption{Same as \cref{fig:fid_fit_nuisance}, but instead corresponding to the run of \cref{fig:data_fit_FP} (right panel).}
    \label{fig:data_fit_FP_nuisance_2}
\end{figure}

\end{document}